\documentclass[aps,prx,twocolumn,a4paper,10pt,nofootinbib,showkeys,longbibliography,nobalancelastpage]{revtex4-2}
\usepackage[utf8]{inputenc}
\usepackage[T1]{fontenc}
\usepackage{lmodern}
\usepackage[english]{babel}
\usepackage{xcolor}
\usepackage[intlimits]{amsmath}
\usepackage{amssymb,amsthm}
\usepackage{mathtools}
\usepackage{graphicx}
\usepackage{textcomp}
\usepackage{enumerate}
\definecolor{green}{HTML}{00643A}
\usepackage[colorlinks=true,allcolors=green,breaklinks=true]{hyperref}
\usepackage{booktabs}
\usepackage{multirow}
\usepackage[]{units}
\usepackage{bm}
\usepackage{url}

\newcommand{\cb}{\text{b}}
\newcommand{\cc}{\text{c}}
\newcommand{\ct}{\text{t}}
\newcommand{\co}{\text{o}}

\newcommand{\kT}{k_\text{B}T}
\newcommand{\cf}{\emph{cf.~}}

\newcommand{\eg}{\emph{e.g.~}}
\newcommand{\ie}{\emph{i.e.}}

\newcommand{\Mc}{\text{M}_\text{c}}
\newcommand{\Mo}{\text{M}_\text{o}}
\newcommand{\Sc}{\text{S}_\text{c}}
\newcommand{\So}{\text{S}_\text{o}}

\begin{document}

\title{Designing bistable nanostructures for target behavior}
\author{Andreas Ehrmann}
\author{Marija Krsti\'c}
\author{Sahar Samadzadeh}
\author{Carl P. Goodrich}
\email{carl.goodrich@ist.ac.at}
\affiliation{Institute of Science and Technology Austria (ISTA), Am Campus 1, 3400 Klosterneuburg, Austria}
\date{\today}

\begin{abstract}

Many biological machines function through controlled conformational transitions, yet designing synthetic nanostructures with prescribed dynamical behavior remains a major challenge. Here, we develop a modular inverse-design framework for bistable nanostructures whose function is controlled by an energy profile along a geometric reaction coordinate. Inspired by proteins with rigid domains connected by flexible hinges, we introduce a hinge-arm paradigm in which a small bistable hinge controls the energetics of a conformational transition, while rigid arms map this transition onto the separation between external binding sites. Specifically, we ask which features of a target energy profile can be programmed under different design constraints. We find that the energy barriers and the binding-site separations in the two metastable states can be readily designed, while controlling the location of the transition state or the full shape of the energy profile requires additional design freedom. Using a differentiable design framework, we find that some optimized solutions are numerically inexact but still display the functional behavior for which the target profile was selected, emphasizing the importance of function-based evaluation criteria. These results establish a practical hierarchy of designability for bistable nanostructures and provide a route toward synthetic nanomachines that couple conformational transitions to target behavior. 

\end{abstract}

\keywords{bistable and functional nanostructures, conformational transitions, inverse design, bio-inspired nanomachines.}

\maketitle

\section{Introduction}
\begin{figure}[t]
    \includegraphics[width=0.49\textwidth]{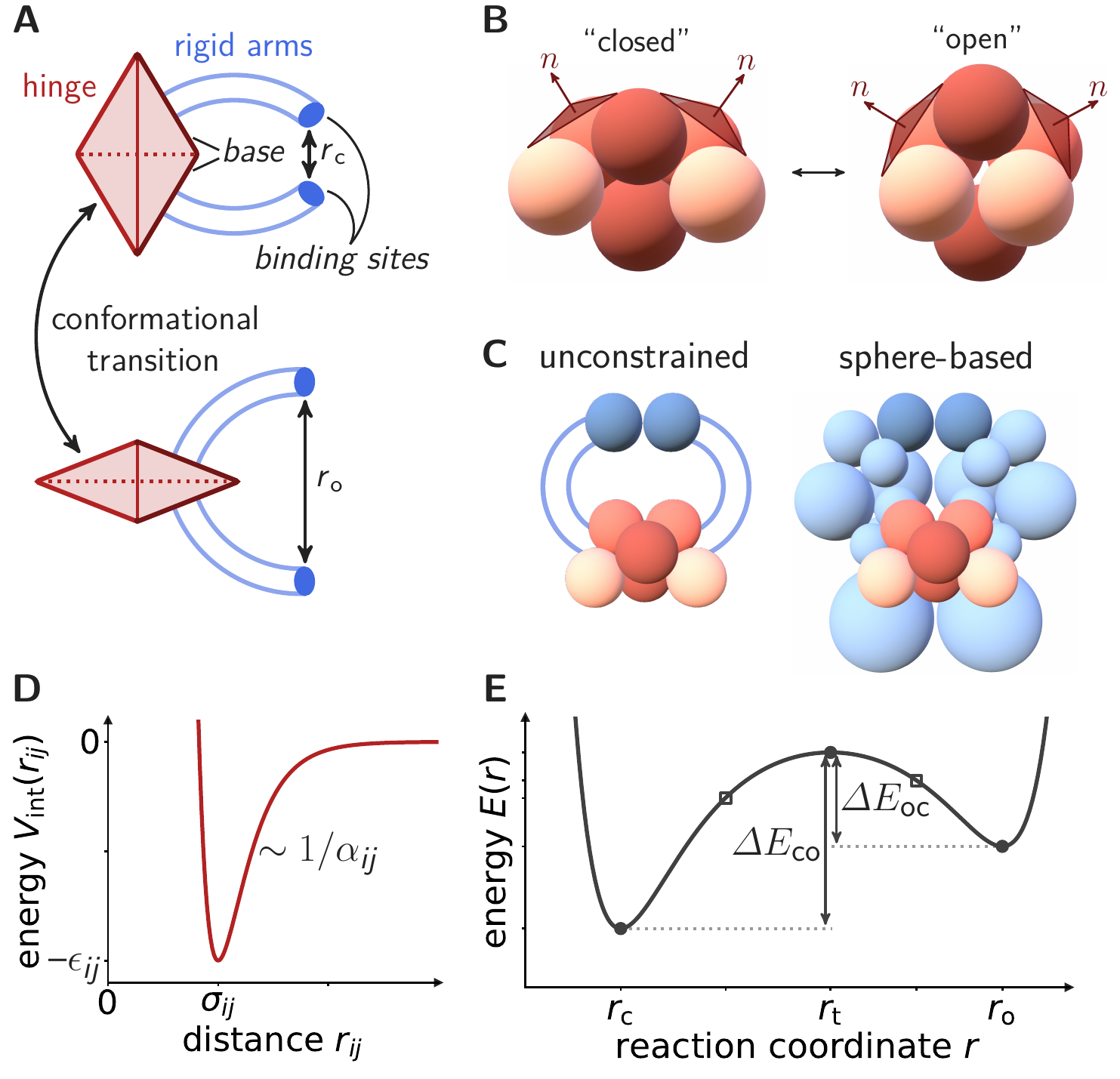}
    \caption{Designing bistable nanostructures for target behavior. \textsf{(A)} The nanostructures consist of a central ``hinge'' region (red) transitioning between two configurations and rigid ``arms'' (light blue) attached to the hinge base (dark red). The arm ends have binding sites (dark blue) to mediate interactions with other nanostructures. The structures are symmetric with respect to the dotted horizontal line. \textsf{(B)} Hinge modeled by six spherical particles, which can be arranged in a ``closed'' polyhedral conformation (left image) with distance $r_\cc$ or an ``open'' octahedral conformation (right image) with distance $r_\co$. The normal vector $n$ to the hinge base, on which the symmetric arms are attached, differs for both structures. \textsf{(C)} Two different arm models. In the unconstrained arm model (Scenarios 1 and 3), we let the arm have an arbitrary shape and consider only the binding sites at a reference location $R_\text{ref}$ with mapping function to the hinge base. In the sphere-based arm model (Scenarios 2 and 4), we model helical arms with spheres of variable diameters. Note that the perspective is rotated compared to \textsf{(B)} so that the normal vectors where the arms are attached point downwards. \textsf{(D)} Interaction potential between two spheres $i$ and $j$, modeled by a Morse potential with interaction strength $\epsilon_{ij}$ and interaction range $1/\alpha_{ij}$ (details in Methods). The force reaches its maximum value at $\ln 2/\alpha_{ij}$. \textsf{(E)} Associated to the conformational change is an energy profile $E(r)$ as function of a reaction coordinate $r$ defined as the distance between the binding sites. Main characteristics are the meta-stable closed and open configurations $r_\cc$ and $r_\co$, the transition state $r_\ct$ (filled circles), the energy barriers $\varDelta E_\text{co}$ and $\varDelta E_\text{oc}$, as well as the overall shape, which we approximate through energies $E_1$ and $E_2$ at specific intermediate values of $r$ (open squares).}
    \label{fig:setup}
\end{figure}

From enzymes to transport proteins to ion channels, countless biological objects rely on controlled changes in their structure while performing a functional task~\cite{phil12,nels13}. 
In many cases, such as proteins like calmodulin and hemoglobin or enzymes like glucokinase, phosphoglycerate kinase, and hexokinase, the structure consists of two rigid domains, or ``arms,'' connected by a flexible or multistable hinge~\cite{pdb101,peru78,fisc11}, see schematic illustration in Fig.~\ref{fig:setup}\textsf{A}. In this way, large-scale motion of binding sites or other active regions located on the rigid arms is controlled by a structural change localized at the hinge~\cite{nels13,zhan19}. 
How can this hinge-arm paradigm be harnessed in \textit{de novo} structures and synthetic nanotechnology to design for specific functional behavior?
To what extent can we program artificial enzymes or catalytically active nanomaterials~\cite{lin14,zhan19}? Despite increasingly rapid advances in numerous strategies for precise design at the nanoscale, including \emph{de novo} protein design~\cite{prae23}, colloidal self-assembly~\cite{roge16,king24,zhu24}, patchy particles~\cite{koeh24}, and DNA origami~\cite{hubl26}, we still lack robust design rules for functionality. 

Here, we focus on the minimal scenario in which the structure interacts with other structures only through two binding sites fixed to the rigid arms. In this case, the only way to \textit{change} the interactions with another structure is to alter the relative positions of these binding sites, which can only be accomplished through a conformational change at the hinge. The kinetics and thermodynamics of the structural change is determined by the internal energy $E$ of the hinge along its reaction coordinate, while the shape and size of the arms determine how the hinge motion affects the separation $r$ between the binding sites. Thus, we can write the internal energy as a function of the separation, $E(r)$; a generic example is illustrated in Fig.~\ref{fig:setup}\textsf{E}. This quantity controls functional behavior and is therefore the central object of interest when seeking to program functionality.

To what extent can $E(r)$ be programmed in practice? 
Using a minimal particle-based model for the hinge, this paper systematically studies the designability of $E(r)$ and shows the conditions under which arbitrary $E(r)$'s can be designed. The primary benefit of the hinge-arm paradigm is that the hinge exclusively controls the energetics, while the arms map the conformation of the hinge to the separation of the binding sites, $r$. We will see that the forward and backward energy barriers, as well as the value of $r$ in the two metastable states, can be programmed relatively easily, while achieving more precise control, \eg over the location of the transition state, is possible but more difficult. These results allow us to classify what features of $E(r)$ can be achieved given the various tunable aspects of the design. 
Finally, as a primary and informative example, we present a particular design that can capture the phenomenology of an ATP-hydrolysis-like mechanism that was previously developed in the context of a dimer with a freely adjustable $E(r)$~\cite{ehrm25}. This example demonstrates the importance of a function-oriented assessment of design success, as we find that some good but numerically imperfect designs nevertheless perform well at the task for which they were designed. These results present, enable, and motivate a concrete design scheme for functional nanostructures, setting the stage for functional design in a wide range of experimental platforms.

\section{A designable hinge-arm model}
We consider a modular approach to designing a bistable nanostructure by first modeling the hinge and then attaching two identical arms. This allows us to design the hinge and arms separately. 

\paragraph*{The hinge.}
While there are many routes for designing the hinge, we consider a simple model composed of six spheres that interact with a Morse potential, see Fig.~\ref{fig:setup}\textsf{D}, with interaction strengths $\epsilon_{ij}$ between particles $i$ and $j$, and interaction ranges $1/\alpha_{ij}$ (see details in Methods). We fix all particle diameters to 1.
Figure~\ref{fig:setup}\textsf{B} shows the six spheres arranged in a polytetrahedron and in an octahedron, and the system can transition between these two conformations by breaking the bond between the two dark red particles and forming a bond between the two light red particles, or vice versa. 
The adjustable design parameters for the hinge are the particle-particle interaction strengths $\epsilon_{ij}$ and, optionally, the inverse interaction ranges $\alpha_{ij}$. 
Previous work has demonstrated how to independently tune the forward and backward energy barriers of such a system~\cite{holm13,good21}, which can be achieved by adjusting the strength of the bonds that form and break in Fig.~\ref{fig:setup}\textsf{B} while imposing high binding strengths between all other bound particles to prevent other transitions. 
We calculate transition pathways between the polytetrahedron and the octahedron states using the doubly-nudged elastic band method~\cite{tryg04}. This also gives an estimate for the transition state, which we refine using eigenvector following~\cite{cerj81,wale04,maur05}.

We also define two triples of hinge particles (indicated by the triangles in Fig.~\ref{fig:setup}\textsf{B}), which each form a ``base'' for attaching the arms. Note the change in the orientation of the vector normal to the base during the structural transition. For the arms we consider below, the binding sites are closer together in the polytetrahedron, which we therefore label as the ``closed'' state, compared to the octahedron, which we label as ``open.''
Importantly, this hinge-arm paradigm is general and can be used with other hinge designs. 
We have tested other transitions for the hinge, including a hinge mechanism between two polytetrahedron conformations, with no noticeable effects (see SI).

\paragraph*{The arms.}
We will consider two distinct models for the arms. 
Importantly, since the functional behavior of the structure is determined by the location of the binding sites at the arm tips, the crucial feature of the arms is this position relative to the location on the switch where the arm is attached. Since we assume that the arms are completely rigid, the details of the arm shape beyond this only affect functionality through steric constraints.
Therefore, the important difference in these two models will not be the internal structure of the arms, but rather how the arms are parameterized, and thus how smoothly and freely the position of the binding site at the arm end can be manipulated. 

In our first model, which we call the ``unconstrained'' design, we let the arm have an arbitrary shape so that we can freely and arbitrarily choose the $(x,y,z)$ coordinates $R_\text{ref}$ of the binding site relative to the base on the hinge (see Fig.~\ref{fig:setup}\textsf{C}). This design has three adjustable parameters, the $x,y,z$ coordinates of $R_\text{ref}$, and can be implemented in practice with any arm shape that results in the same $R_\text{ref}$. Our second model is a ``sphere-based'' design, where a helical spiral of $N$ spheres is attached at the base (see SI for details). Here, the adjustable parameters are the $N$ diameters of the spheres: growing or shrinking a sphere will bend the arm and thus affect $R_\text{ref}$. Note that while this is actually overparametrized, deriving $R_\text{ref}$ from spheres adds nontrivial constraints, \eg due to steric interactions between the two arms. Thus, this sphere-based model allows us to explore the consequence of potential experimental design constraints. 

\paragraph*{Designing $E(r)$.}
This hinge-arm model results in a well-defined energy profile, $E(r)$, similar to the example shown in Fig.~\ref{fig:setup}\textsf{E}. Our goal is to match this to a particular desired target profile. We do this by identifying key features of the target profile and constructing a loss function that quantifies how close the observed features are to the target. We inverse-design by using methods of Automatic Differentiation (AD)~\cite{bayd18,rume86,weng64} to calculate the gradient of the loss. We then use the JAXopt~\cite{blon22} implementation of the L-BFGS-B algorithm to iteratively minimize the loss by adjusting the relevant model parameters (Methods). 

Which features of $E(r)$ should be considered, how close they need to be to the target for the optimization to be considered successful, and which model parameters are adjusted all depend on the context. 
Potential relevant features of $E(r)$ include: the binding-site separation in the closed state, $r_\cc$, the open state, $r_\co$, and the transition state, $r_\ct$, the forward and backward energy barriers, $\varDelta E_\text{co}$, and $\varDelta E_\text{oc}$, respectively, and the energy at various intermediate values of $r$ along the energy profile. We combine these hierarchically into four distinct loss functions: 
\begin{align}
\begin{split}
    \mathcal{L}_1 &= (\varDelta E_\text{co} - \varDelta E_\text{co}^*)^2 + (\varDelta E_\text{oc} - \varDelta E_\text{oc}^*)^2, \\
    \mathcal{L}_2 &= \mathcal{L}_1 + (r_\cc - r_\cc^*)^2 + (r_\co - r_\co^*)^2, \\
    \mathcal{L}_3 &= \mathcal{L}_2 + (r_\ct - r_\ct^*)^2, \\
    \mathcal{L}_4 &= \mathcal{L}_3 + (E_1 - E_1^*)^2 + (E_2 - E_2^*)^2,
\end{split}\label{eq:loss_functions}
\end{align}
where $E_1$ and $E_2$ are the energies at specific intermediate values of $r$, and quantities indicated with a ``$^*$'' are the desired values taken from the target profile. We also add additional physical constraints to the loss functions, see Methods, for example in the sphere-based arm model to prevent the spheres from ever overlapping.

Using these four loss functions, we will attempt to inverse design two target energy profiles that were developed previously~\cite{ehrm25} as part of an ATP-like energy delivery mechanism. For our present purposes, they are simply reasonable example profiles that one may wish for. We refer to the first profile (dashed black line in Fig.~\ref{fig:energy}) as the ``easy'' profile because it is roughly symmetric: the transition state is roughly half way between the open and closed states, and the forward and backward energy barriers are comparable. On the other hand, the second profile (dashed blue line in Fig.~\ref{fig:energy}) is ``hard'' because the transition state is very close to the closed state and the backwards energy barrier is significantly larger than the forwards barrier, making for a more challenging target.

Finally, to explore the ability to inverse-design these nontrivial energy profiles in different contexts, we will consider four distinct design scenarios. In Scenario 1, we consider the unconstrained arm model with a hinge where the inverse interaction ranges, $\alpha_{ij}$, are fixed. Scenario 2 is identical to Scenario 1 except that we use the sphere-based model for the arms with $N=11$ spheres per arm. The point here is to explore the role of additional constraints and design complications that make the design process more realistic. Scenario 3 returns to the unconstrained arm model but includes the inverse interaction ranges $\alpha_{ij}$ as design variables. Finally, Scenario 4 combines variable $\alpha_{ij}$ with the sphere-based model for the arms. In the next section, we will explore when and why these four scenarios are able to successfully design nontrivial energy profiles. 

\begin{figure}[t]
    \includegraphics[width=0.49\textwidth]{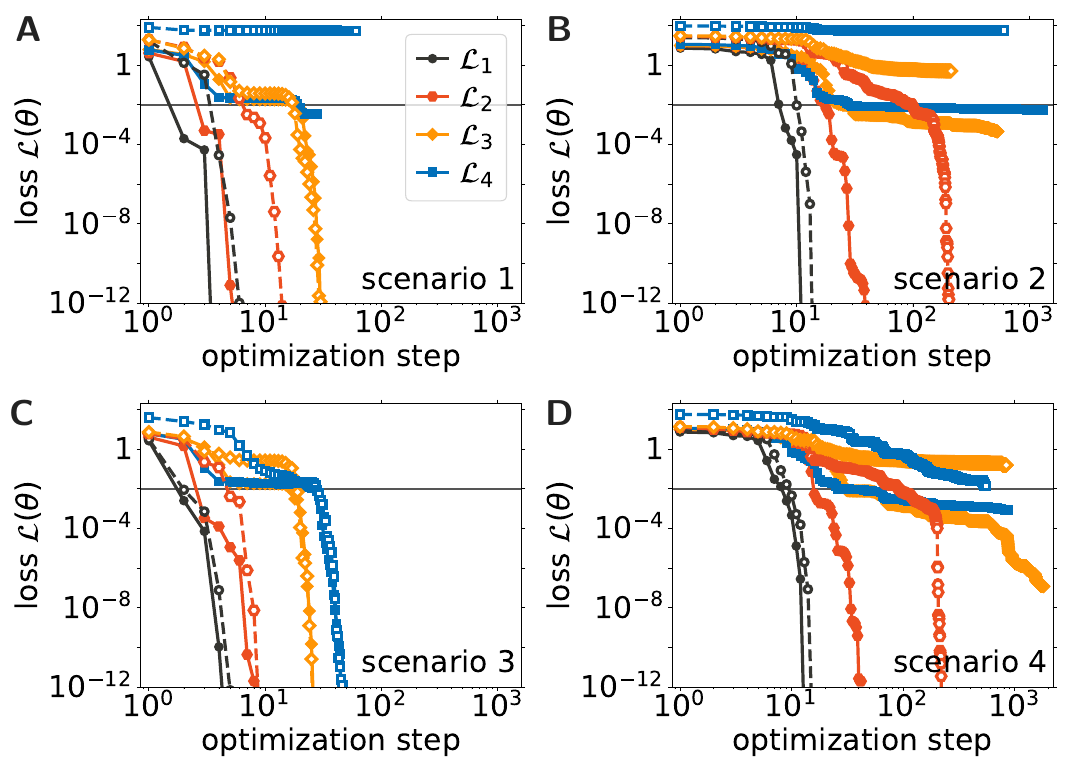}
    \caption{Hierarchical designability optimization outcomes with increasingly complex loss functions $\mathcal{L}_1$ to $\mathcal{L}_4$ from Eq.~\eqref{eq:loss_functions} (as described in the main text) for all four different design scenarios. Solid lines with solid symbols are for the simpler energy profile in Fig.~\ref{fig:energy} and dashed lines with open symbols for the harder profile. Note that the reason for the different convergence behavior for the simple loss functions lies in the additional constraints in the sphere-based arm model. The horizontal line at $\mathcal{L}=10^{-2}$ indicates solutions that are ``good enough / functional'' by passing the eye test (see Fig.~\ref{fig:energy}).}
    \label{fig:designability}
\end{figure}

\begin{figure}[t]
    \includegraphics[width=0.49\textwidth]{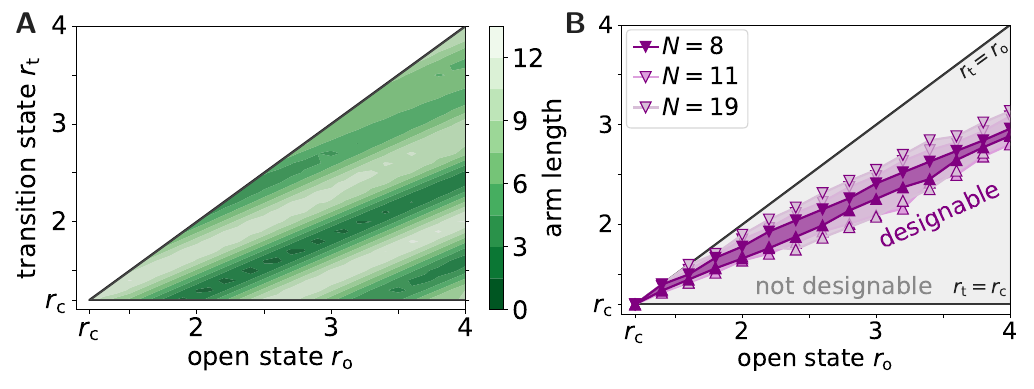}
    \caption{Design freedom in the location of the transition state in two design scenarios for modeling rigid arms. The locations of the meta-stable states are fixed at $r_\cc=1.2$ and $r_\co$. \textsf{(A)} For the unconstrained arm model (Scenario 1), all values of the transition state location $r_\ct$ between $r_\cc$ and $r_\co$ can be achieved with high accuracy, \ie, $\mathcal{L}_3^\mathrm{final}<10^{-12}$. The color indicates the length of the arm, defined as distance from the binding site to the hinge center, in the optimal design. \textsf{(B)} Control over the location of the transition state is limited in Scenario 2 when varying the colloid diameters in the sphere-based arm model. We find that the transition state is `nearly designable' (defined as $\mathcal{L}_3^\mathrm{final}<10^{-2}$) in the purple region, which increases for increasing numbers of colloids in the arms $N=8,11,19$, and is not designable in the gray region. The slope of the design bands is determined by the hinge design as shown in the SI. The comparison of these design scenarios informs how experimental constraints can impact functional design.}
    \label{fig:design_rt}
\end{figure}

\section{Inverse-designing functional nanostructures}
Figure~\ref{fig:designability} shows the optimization results for the four design scenarios (\textsf{A-D}), the four loss functions (colors) and the two target profiles (open and closed symbols). In all cases, $\mathcal{L}_1$ is reduced below $10^{-12}$. This is consistent with previous work~\cite{good21} that demonstrates the ability to simultaneously and accurately design forward and backward energy barriers with high precision. While this result was expected and included for completeness, $\mathcal{L}_2$ also drops below $10^{-12}$ in all cases. In all scenarios, the positions of both minima, $r_c$ and $r_o$, can be readily tuned with high accuracy.

The quality of the designs become more nuanced when including the position of the transition state in the objective, $\mathcal{L}_3$. Figure~\ref{fig:designability}\textsf{A} shows that $\mathcal{L}_3$ vanishes for Scenario 1, so that all five relevant features are precisely matched in both the easy and hard profiles. To show that these results are general, enabling the design of any combination of $r_\cc^*$, $r_\co^*$, and $r_\ct^*$,  we systematically change the target values of these features and find that we are always able to minimize $\mathcal{L}_3$ below $10^{-12}$. This is consistent with the fact that the arm design consists of three parameters: the coordinates of $R_\text{ref}$. For fixed $r_\cc^*=1.2$, Fig.~\ref{fig:design_rt}\textsf{A} shows the length of the optimized arm for a range of $r_\co^* > r_\cc^*$ and $r_\ct^*$ between $r_\cc^*$ and $r_\co^*$. Data for other $r_\cc^*$ is shown in the SI, together with an explanation that the slope of the design bands is determined by the hinge design.

\begin{figure}[t]
    \includegraphics[width=0.49\textwidth]{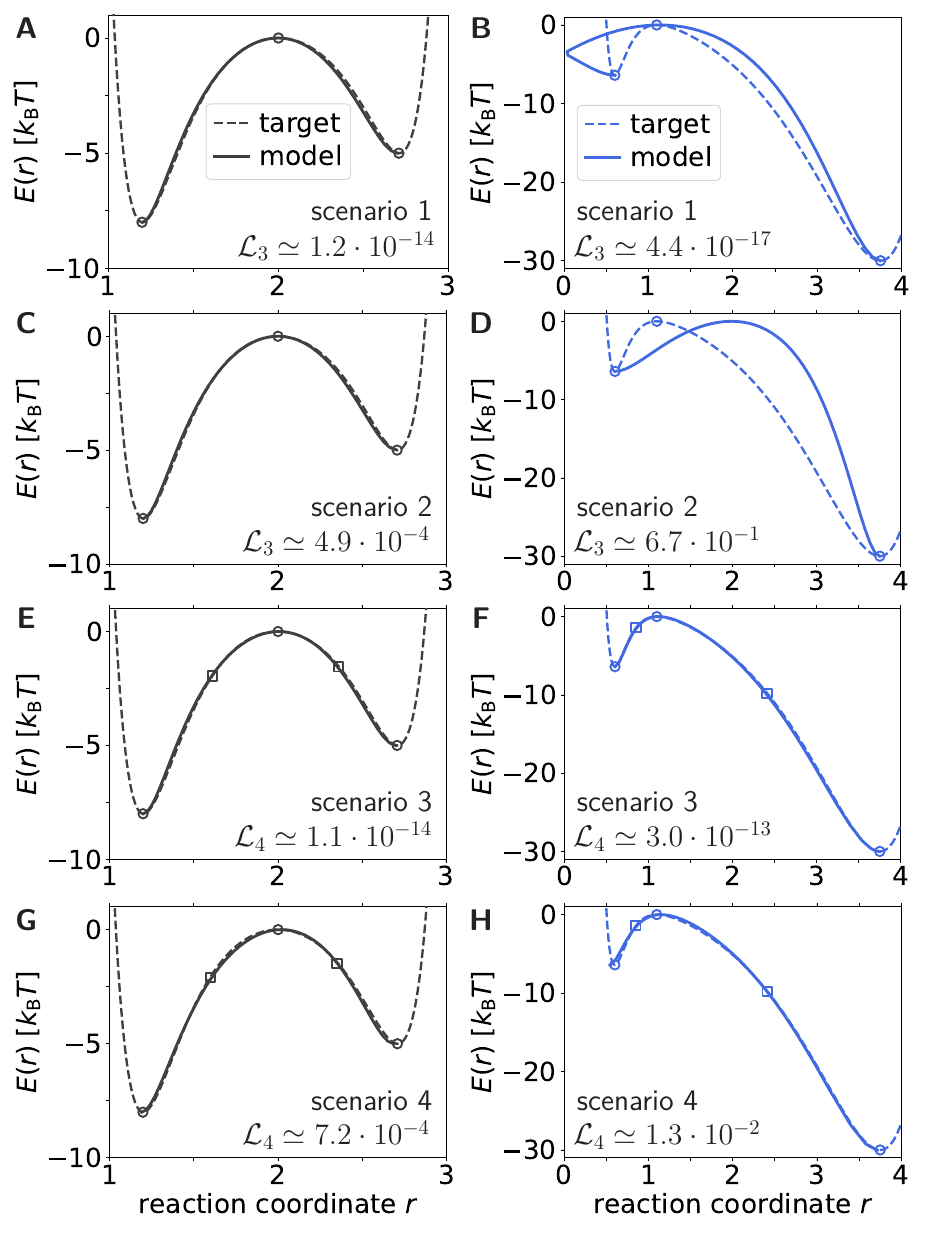}
    \caption{Matching target energy profiles in different design scenarios for functional nanostructures. We investigate four design scenarios, which are defined in Table~\ref{tab:designability}, for two target energy profiles (dashed black and blue curves). In each scenario, we perform an optimization of the respective loss function in Eq.~\eqref{eq:loss_functions}, with target values indicated by open symbols, over the respective model parameters as described in the main text. We use the qualitative agreement of the resulting energy profiles (solid curves) with the target profiles (dashed curves) in each row to infer the designability ranking $\checkmark/\times$ in Table~\ref{tab:designability}. The design scenarios with variable interaction ranges show an excellent (Scenario 3) or good (Scenario 4) agreement with the target energy profiles. This demonstrates that we have precise or good control over the energy profiles with our computational framework. We state the final loss values after the optimization has converged (details in SI).}
    \label{fig:energy}
\end{figure}

To better understand the quality of these designs, we plot the actual energy profiles produced by the optimization and visually compare them to the target. For the case of targeting $\mathcal{L}_3$ with Scenario 1, this is shown for the two example target profiles in Fig.~\ref{fig:energy}\textsf{A-B}. Visual inspection of the easy profile, Fig.~\ref{fig:energy}\textsf{A}, shows excellent agreement even in the intermediate regions where $E(r)$ does not enter the loss function. However, visual inspection of the hard profile, Fig.~\ref{fig:energy}\textsf{B}, shows significant deviations in all intermediate regions. During the transition, the two arms actually ``pass through'' each other, resulting in $r$ momentarily becoming zero (see SI for discussion). 
This is not inconsistent with $\mathcal{L}_3<10^{-12}$ because $\mathcal{L}_3$ does not depend on what happens in these intermediate regions; however, it implies that this solution is likely unphysical and highlights the importance of including additional physical constraints to prevent this.

Such constraints are incorporated into Scenario 2, which again is identical to Scenario 1 except that we use the sphere-based model, preventing the ``passing through'' behavior observed in Fig.~\ref{fig:energy}\textsf{B}. Figure~\ref{fig:designability}\textsf{B} shows that optimization is indeed more difficult with the sphere-based model. 
While visual inspection of the optimized easy profile (Fig.~\ref{fig:energy}\textsf{C}) appears quite good, with only minor intermediate deviations from the target profile, $\mathcal{L}_3$ does not drop below $10^{-4}$. Furthermore, optimization of the hard profile is clearly unsuccessful: $\mathcal{L}_3$ never drops below 0.6, and the position of the transition state, $r_\ct$, noticeably deviates from its target (Fig.~\ref{fig:energy}\textsf{D}). Apparently, if the arms are not allowed to ``pass through'' each other, the ability to design for $r_\ct$ becomes uncertain. 

While neither objective is perfectly satisfied, there is a clear qualitative difference between the easy and hard profiles attained in Fig.~\ref{fig:energy}\textsf{C-D}. When does a target design fail, as in Fig.~\ref{fig:energy}\textsf{D}, and when is it nearly successful, as in Fig.~\ref{fig:energy}\textsf{C}? We systematically address this question in Fig.~\ref{fig:design_rt}\textsf{B}, which shows which values of $r_\ct$ can be `nearly designed' (defined by $\mathcal{L}^\mathrm{final}_3 < 10^{-2}$) while keeping $r_\cc$ and $r_\co$ fixed to their targets. As one might expect, this region increases for larger $N$, as this gives the system access to more degrees of freedom and longer arms. 

In order to better design the full energy profile, we turn to Scenario 3, where the unconstrained arm model is employed while including the interaction ranges $1/\alpha_{ij}$ as design variables. This provides additional degrees of freedom, located within the hinge, that can be used to refine the value of $r$ throughout the transition. As with Scenario 1, Scenario 3 can easily reduce $\mathcal{L}_3$ below $10^{-12}$ (Fig.~\ref{fig:designability}\textsf{C}). However, we now see that $\mathcal{L}_4$, which includes values of the energy at intermediate points along the profile, can also drop below $10^{-12}$ for both the easy and hard profiles. Note that this was not possible in Scenario 1 (Fig.~\ref{fig:designability}\textsf{A}) because the unconstrained arm model simply lacks the necessary degrees of freedom to independently control five points along the profile. Visual inspection of the resulting profiles, Fig.~\ref{fig:energy}\textsf{E-F}, confirms that these designs show excellent agreement with the target profiles, and the arms do not ``pass through'' each other as they did in Scenario 1 because the location of $r_\ct$ is controlled in part by the $\alpha_{ij}$ variables rather than just the arms. 

Finally, Scenario 4 combines variable $\alpha_{ij}$ with the sphere-based model for the arms. Figure~\ref{fig:designability}\textsf{D} shows that neither $\mathcal{L}_3$ nor $\mathcal{L}_4$ drop below $10^{-12}$ for either of the target profiles. However, like with Fig.~\ref{fig:energy}\textsf{C}, visual inspection of the optimized profiles shows very close agreement. While minor quantitative differences keep the loss as large as $10^{-2}$, they appear successful to the eye as did the design for the easy profile in Scenario 2. Importantly, unlike the clear failure of Fig.~\ref{fig:energy}\textsf{D}, the discrepancies are not due to the constraints of the spheres. Rather, the problem stems from the indirect parameterization of the arms, which makes precise fine-tuning of the final $R_\mathrm{ref}$ difficult (SI). 
The results in Fig.~\ref{fig:design_rt}\textsf{B} demonstrate that increasing the number of particles in the arms improves the range over which we achieve nearly successful designs, while also improving their quality (not shown). 

We are now faced with an interesting question: are the designs presented in Fig.~\ref{fig:energy}\textsf{C}, \textsf{G}, and \textsf{H} successful? 
From a numerical perspective, the answer is clearly no as the final losses, which range from $\sim 10^{-4}$ to $\sim 10^{-2}$ do not approach numerical zero. However, if these designs were achieved in an experiment, one would hope that the small deviations from the target are unimportant. This is not to say that these designs should be deemed successful at the same level as other designs where the loss drops below $10^{-12}$, but it highlights that the evaluation of success depends on context. Are the deviations in a design of this type important for the underlying behavior for which the target profile is intended to achieve? Answering this question can only be made on a case-by-case basis in the context of larger design objectives beyond the objectives discussed here. In the next section, we will investigate how this question can be addressed more robustly.

\begin{figure}[t]
    \includegraphics[width=0.49\textwidth]{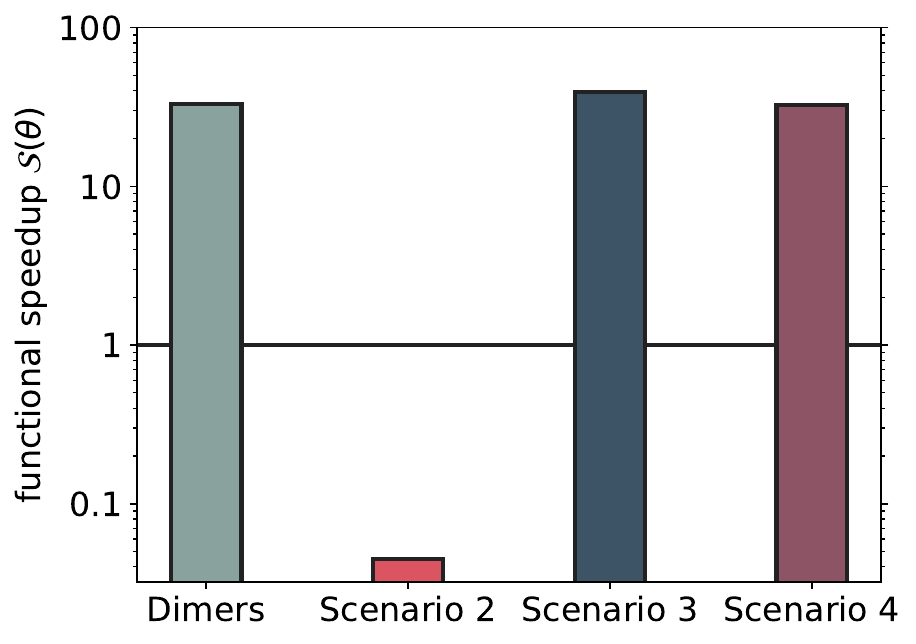}
    \caption{Functional speedup of the energy-delivery mechanism for different systems. The functional speedup defined in Eq.~\eqref{eq:speedup} of bistable nanostructures with design Scenarios 3 and 4 is very close to the value obtained for bistable dimers. The energy profiles of Machine and Source nanostructures are modeled using the solid lines in Fig.~\ref{fig:energy}\textsf{C-H} for the different scenarios and the dimers are modeled using the target profiles (dashed lines in Fig.~\ref{fig:energy}). The speedup is calculated for an interaction strength of $\epsilon=10\,\kT$ for all systems. The speedup values are 33.17 (dimers), 0.04 (Scenario 2), 39.14 (Scenario 3), and 32.69 (Scenario 4). The model parameters are listed in Table~\ref{tab:params}.}
    \label{fig:speedup}
\end{figure}

\section{Functional determination of success}
The two target profiles we have used as primary demonstrations in Figs.~\ref{fig:designability} and \ref{fig:energy} were developed as part of an ATP-like energy delivery mechanism~\cite{ehrm25}. Energy delivery is achieved by considering two 2-state structures, one with the easy profile and one with the hard profile, where both structures begin in their closed states. The target profiles were themselves designed so that when these structures encounter each other, they form a multi-state complex where both structures can open together. In other words, these profiles enable a coupled reaction, the result of which is that some of the stored potential energy in the structure with the hard profile is transferred to the structure with the easy profile. 

Understanding the functional purpose of the target profiles is important because it provides a more practical means to evaluate the designs presented in Fig.~\ref{fig:energy}: a design is considered successful if it achieves the larger design objective of enabling energy delivery. To test this, we measure the functional speedup of the reaction, which is a proxy for the effectiveness of the design and was used in Ref.~\cite{ehrm25} to design the profiles that we use here as targets (see Methods for details).

Figure~\ref{fig:speedup} compares the functional speedup, $\mathcal{S}$, for the energy profiles obtained in Scenarios 2, 3, and 4 (shown in Fig.~\ref{fig:energy}) to that of the target profiles. The results of Scenario 1 exhibit the unphysical ``passing through'' behavior discussed above, and are thus not considered here. The value of the functional speedup directly determines the effectiveness of the energy-delivery mechanism, with $\mathcal{S}>1$ necessary to drive non-equilibrium behavior. 

The results shown in Fig.~\ref{fig:speedup} are very clear. The profiles attained in Scenario 2 do not work: the mismatch in $r_\ct$ for the hard profile (Fig.~\ref{fig:energy}\textsf{D}) disrupts the coupled reaction, leading to a functional speedup that is not only less than 1, but nearly three orders of magnitude less than that of the target profiles. However, the profiles of Scenarios 3 and 4 perform excellently, with functional speedups very close to the target. Scenario 3 actually slightly outperforms the target. Therefore, we conclude that the designs of Scenarios 3 and 4 are successful regardless of the numerical precision of the optimization, while those of Scenarios 1 and 2 are unsuccessful.

\section{Discussion}

We have shown how bistable, two-state nanostructures can be designed to realize target energy profiles, and we have identified which features of those profiles are accessible under different design constraints.
Inspired by various proteins and enzymes, we introduce a hinge-arm paradigm where the structural transition is localized to a small region in the structure, and then rigid arms translate this reorganization into a change in the separation, $r$, between two binding sites. This separation provides a clear means for other objects to interact with the nanostructure based on its structural state, and therefore presents a convenient, measurable, and important reaction coordinate for the structural transition. As demonstrated in other work~\cite{ehrm25}, the shape of $E(r)$ determines the structure's function, with key features including the energy barriers, the value of $r$ at the closed state, open state, and transition state, and finally the energy at intermediate $r$. 

This hinge-arm paradigm leads to a clear, modular, hierarchical, and adaptable strategy for inverse design. First, the inverse-design of barrier heights has already been demonstrated for a switch composed of a small cluster of spheres with specific binding energies~\cite{good21}, and achieving such behavior in other systems is an active research area~\cite{prae23,hubl26,meli26}. Next, rigid arms control the $x,y,z$ position of the binding site relative to the position and orientation of a region of the switch where it is attached, called the base. Optimization is performed by combining methods of Automatic Differentiation~\cite{bayd18,rume86,weng64} with standard gradient-based minimization routines. The three degrees of freedom that define the binding sites for symmetric rigid arms enable arbitrary and simultaneous control of the binding-site separation at the closed state, the open state, and the transition state. However, some of these solutions lead to very large arms and can even involve arms that ``pass through'' each other, \eg see Fig.~\ref{fig:energy}\textsf{B}. 

We therefore also consider an arm model composed of a helical assembly of spheres, giving the arms a well-defined and controllable shape that allows us to impose overlap constraints in the optimization process. These overlap constraints prevent the simultaneous control of $r$ at the closed state, the open state, and the transition state, except for special cases, see Fig.~\ref{fig:design_rt}\textsf{B}, meaning that rigid symmetric arms alone are not sufficient if specifying all three separations is important. If they are, we achieve excellent designs by returning to the hinge and allowing the interaction ranges of the constituent particles to vary. This changes the configuration of the hinge that corresponds to the transition state, and therefore allows more fine-tuned control even with the sphere-based arm model. Furthermore, this allows for the design of intermediate values $E(r)$ throughout the energy profile.

Importantly, the optimization process for some designs leads to loss values that are small but numerically non-zero, and it is not obvious if they should be considered successful. We take a functional approach to determining success. The target profiles themselves do not have intrinsic value, but rather are desired for a particular reason, \eg they lead to some desired functional behavior. Therefore, the relevant question is whether the profiles attained exhibit or enable that behavior. This functional assessment is important because, even if a model nanomachine is designed with an energy profile that is exactly as desired, actual experimental realizations will not match this perfectly. Therefore, a functionality-based understanding of success is necessary before translating such hinge-arm designs to the lab. 

These results demonstrate and evaluate the design potential of the hinge-arm paradigm through a range of model assumptions and design objectives, giving crucial insight into when, why, and how such a paradigm should be pursued. We have intentionally kept our analysis as general as possible without committing to any particular experimental platform. Nevertheless, for practitioners seeking to design bistable nanostructures with specific functional behavior, the hierarchy presented in Table~\ref{tab:designability} summarizes our results to provide a practical roadmap. This allows one to read off which design features are achievable under different experimental constraints, and which design variables are necessary to achieve them. A practical workflow begins by identifying a target energy profile (or at least the key features) that is expected to enable the desired behavior. One can then consult Table~\ref{tab:designability} to determine whether the available experimental control is sufficient to achieve that target. For example, if the location of the transition state is critical for function, then controlling the shape of the arms is likely insufficient unless the issue presented in Fig.~\ref{fig:energy}\textsf{B} can be solved. Importantly, a successful design need not be numerically perfect, and the functional validation step is essential to assess the experimental viability of a design.

\begingroup
\setlength{\tabcolsep}{3pt}
\renewcommand{\arraystretch}{1.5}
\begin{table*}[tb]
    \begin{tabular}{ccc}
        \toprule
        \multicolumn{3}{c}{design strategy} \\
        scenario & variable interaction ranges & arm model \\  
        \midrule
        1 & $\times$ & \includegraphics[height=0.021\textwidth]{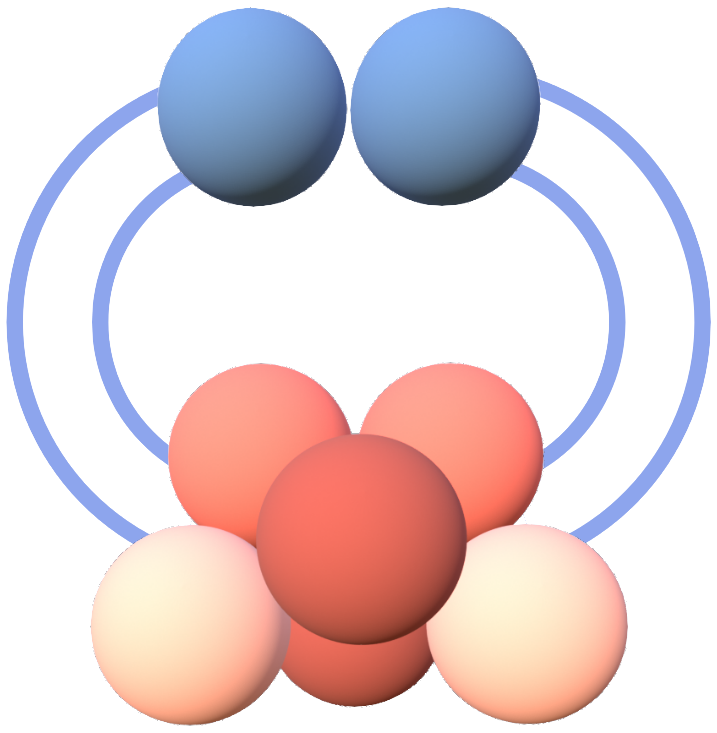} \\
        2 & $\times$ & \includegraphics[height=0.021\textwidth]{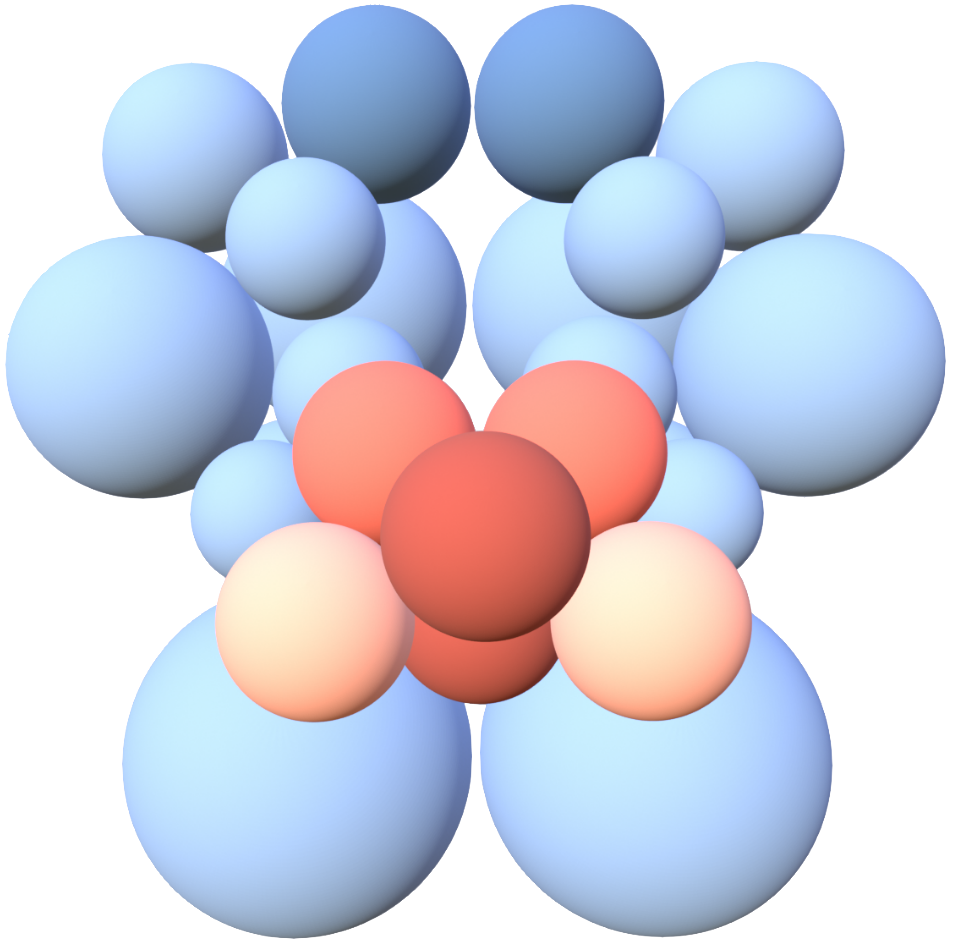} \\
        3 & \checkmark & \includegraphics[height=0.021\textwidth]{models1n} \\
        4 & \checkmark & \includegraphics[height=0.021\textwidth]{models2n} \\
        \bottomrule
    \end{tabular}
    \begin{tabular}{cccc}
        \toprule
        \multicolumn{4}{c}{design objectives} \\
        energy barriers & locations minima & location transition state & shape \\
        \midrule
        \checkmark & \checkmark & \checkmark\footnote{See Figure~\ref{fig:design_rt}\textsf{A}.} & $\times$ \\
        \checkmark & \checkmark & $\times$\footnote{Only the small region in Figure~\ref{fig:design_rt}\textsf{B} is nearly designable.} & $\times$ \\
        \checkmark & \checkmark & \checkmark & \checkmark \\
        \checkmark & \checkmark & $\approx$ & $\approx$ \\
        \bottomrule
    \end{tabular}
    \caption{Hierarchical designability ranking of the four different design scenarios for functional nanostructures. If one has control over the interaction ranges $1/\alpha_{ij}$ in the hinge, in addition to the interaction strengths $\epsilon_{ij}$, this is sufficient to match the full shape of a target energy profile in addition to matching the location $r_\ct$ of the transition state as well as the locations of both minima $r_\cc$, $r_\co$ and the energy barriers $\varDelta E_{\cc\co}$, $\varDelta E_{\co\cc}$. The target energy profiles are matched precisely in Scenario 3 (\checkmark, $\mathcal{L} < 10^{-12}$, unconstrained arm model) or nearly successfully in Scenario 4 ($\approx$, $\mathcal{L} < 10^{-2}$, sphere-based arm model), whereas Scenarios 1 and 2 do not have enough degrees of freedom to address all hierarchical design objectives.}
    \label{tab:designability}
\end{table*}
\endgroup

In both arm models that we consider, we assume that the arms are perfectly rigid, meaning that the binding-site separation is completely determined by the configuration of the hinge. Real nanostructures will exhibit some flexibility, which can affect the functional behavior. For example, if the hinge is in a high-energy state with a binding-site separation near $r_\mathrm{t}$, it is possible to lower the energy while keeping $r$ fixed by elastically deforming the arms to allow the hinge to partially relax. The impact of this depends on the flexibility of the arms and the functional requirements of the nanostructure, and should be evaluated on a case-by-case basis. 

By separating the energetic control of a bistable transition from the geometric control of how that transition is presented to the outside world, the hinge-arm paradigm and the results presented here provide a modular route toward synthetic nanostructures with targeted dynamical behavior. The next step is to integrate this design framework with the additional constraints required by specific experimental systems. In particular, platform-specific models will be important for understanding and designing interactions with other structures, including attachment geometries and steric constraints. By clarifying which features of a bistable energy landscape are designable, which require additional degrees of freedom, and when functional assessment is important, our results provide a foundation for the rational design of synthetic nanomachines that couple conformational transitions to useful behavior.

\section{Materials and methods}

\subsection{Model details}

The interaction potential of two spheres $i$ and $j$ with diameters $D_i$ and $D_j$ at distance $r_{ij}$ is given by a Morse potential
\begin{equation}
    V_\text{int}(r_{ij}) = \epsilon_{ij}\left(e^{-2\alpha_{ij}(r_{ij} - \sigma_{ij})} - 2e^{-\alpha_{ij}(r_{ij} - \sigma_{ij})}\right),
\end{equation}
with equilibrium distance $\sigma_{ij} \equiv (D_i + D_j) / 2$, interaction strength $\epsilon_{ij}$, and interaction range $1/\alpha_{ij}$. The force $-\nabla V_\text{int}$ reaches its maximum value at $\ln 2/\alpha_{ij}$.
We define the total energy of a colloidal nanostructure as sum of pairwise Morse potentials between all bound spheres
\begin{equation}\label{eq:Etot}
    E(r) = \sum_{i\neq j} V_\text{int}(r_{ij}).
\end{equation}
For simplicity we assume that different nanostructures solely interact through their point-like binding sites, modeled via Morse potentials with $\epsilon$ and $\alpha$. 

The optimized parameters used to model bistable nanostructures in Scenario 3 for the unconstrained arm model with variable interaction strengths and interaction ranges in the hinge are listed in Table~\ref{tab:params}.
\begin{table}[t]
 \begin{tabular}{cccccccc}
  \toprule
   & $\epsilon_\cc$ & $\epsilon_\co$ & $\epsilon_\cb$ & $\alpha_\cc$ & $\alpha_\co$ & $\alpha_\cb$ & $R_\text{ref}$ \\
  \midrule
  Machine & $11.23$ & $7.54$ & $100$ & $8.16$ & $8.37$ & $50$ & $(3.35, -2.50, 1.09)$ \\
  Source & $9.09$ & $35.39$ & $100$ & $28.05$ & $4.87$ & $50$ & $(4.96, -4.54, 1.15)$ \\
  \bottomrule
 \end{tabular}
 \caption{List of all parameters used to model bistable nanostructures in design Scenario 3 using the unconstrained arm model with variable interaction strengths and interaction ranges in the hinge. The values correspond to the closed, open, and bound spheres and the coordinates of the binding site for the ``easy'' profile (Machine) and the ``hard'' profile (Source). The model parameters for the other design scenarios can be found in the SI.}
 \label{tab:params}
\end{table}
The corresponding parameters for the other design scenarios can be found in the SI.
The results in this manuscript can be reproduced for any experimental method of choice to design the arms of the nanostructures if they match the given reference locations of the binding sites.

\subsection{Navigating parameter space with differentiable programming}

We have developed a differentiable model to optimize the design of functional nanostructures with respect to physical model parameters. To design the sphere-based model for each set of new parameters during the optimization, we first calculate the energy profile of the hinge using the doubly-nudged elastic band method~\cite{tryg04} and refine the location of the transition state using eigenvector following~\cite{cerj81,wale04,maur05}. We convert the bonds in the structure to springs and run energy minimization on this spring network in order to get the new equilibrium distances. We construct the new arm for these distances and add it to the hinge for all images of the transition pathway. 
For the sphere-based arm model in Scenarios 2 and 4, the loss functions in Eq.~\eqref{eq:loss_functions} additionally include physical constraints. The constraints are given by the sum of all pairwise soft sphere energies between the particles 
\begin{equation}
    E_\text{overlap} = \sum_{i\neq j}\begin{cases}
               5\left(1-\frac{r_{ij}}{\sigma_{ij}}\right)^2, & \text{if } r_{ij}/\sigma_{ij}<1,\\
               0, & \text{if } r_{ij}/\sigma_{ij}\geq1,
    \end{cases}
\end{equation}
to prevent overlaps and the energy of the spring network 
\begin{equation}
    E_\text{position} = \sum_{i\neq j} \frac{1}{2}\left(r_{ij} - \sigma_{ij}\right)^2
\end{equation}
to ensure correct positioning of the particles in the arms. 
In comparison, the unconstrained arm model in Scenarios 1 and 3 does not include any constraints. We calculate the gradient $\nabla_\theta\mathcal{L}$ of the loss function using Automatic Differentiation~\cite{bayd18,rume86,weng64}, and feed this gradient into standard gradient-based optimization routines. We use ScipyMinimize from JAXopt with the standard L-BFGS-B algorithm to adjust the parameters such that all objectives converge to their target values. We terminate the optimization using a tolerance of $10^{-12}$ for a maximum number of $10^5$ iterations. Our work is based on the software packages JAX and JAX-MD~\cite{jax2018github,scho21}, with built-in hardware acceleration and ensemble vectorization.

\subsection{Functional speedup of energy delivery between nanostructures}
To determine if the designed nanostructures indeed lead to a functionally successful target behavior, we study the behavior of a coupled system consisting of two bistable nanostructures matching the target energy profiles discussed above. One of them acts as a ``Source'' of energy through a thermodynamically favorable reaction, $\Sc\to\So$, when transitioning to its open state. The released energy is then harnessed to drive the second structure, acting as a ``Machine'', into its high-energy state, $\Mc\to\Mo$ (thermodynamically unfavorable). We have shown in previous work for a coarse-grained model of dimers~\cite{ehrm25} that the structures can efficiently and effectively transfer energy through this coupled reaction, $\Mc\cdot\Sc \to \Mo\cdot\So$. Now we want to assess whether this still holds for the more realistic realization of nanostructures developed in this work. We define the functional speedup as the ratio of mean first-passage times of the opening transitions for the Machine on its own (equilibrium) and the coupled system (non-equilibrium), 
\begin{equation}\label{eq:speedup}
    \mathcal{S}(\theta) \equiv \frac{\mathcal{T}_\text{eq}(\Mc\to\Mo)}{\mathcal{T}_\text{non-eq}(\Mc\cdot\Sc \to \Mo\cdot\So)},
\end{equation}
as a proxy for the effectiveness of the design. It was used in Ref.~\cite{ehrm25} to design the energy profiles that we use as targets in this work. The value of the functional speedup directly determines the effectiveness of the energy-delivery mechanism, with $\mathcal{S}>1$ necessary to drive non-equilibrium behavior. A detailed discussion of the coupled energy-delivery reaction between both nanostructures can be found in the SI together with an animation of the reaction pathway.

\begin{acknowledgments} 

We thank Maitane Muñoz-Basagoiti for stimulating discussions.
This research was funded in part by the Austrian Science Fund (FWF) [10.55776/PAT8537123].

\end{acknowledgments}

\bibliography{literature}

\appendix

\section{Model details}

\subsection{Designing bistable nanostructures}

The bistable nanostructures used in this work consist of a central ``hinge'' region transitioning between two configurations and rigid ``arms'' that are symmetrically attached to the hinge. The hinge is modeled by six spherical particles with diameter of 1, which can be arranged in a ``closed'' polyhedral conformation or an ``open'' octahedral conformation (see red particles in Fig.~\ref{fig:transition}). In this work, the interaction potential of two spheres $i$ and $j$ with diameters $D_i$ and $D_j$ at distance $r_{ij}$ is given by a Morse potential
\begin{equation}
    V_\text{int}(r_{ij}) = \epsilon_{ij}\left(e^{-2\alpha_{ij}(r_{ij} - \sigma_{ij})} - 2e^{-\alpha_{ij}(r_{ij} - \sigma_{ij})}\right),
\end{equation}
with equilibrium distance $\sigma_{ij} \equiv (D_i + D_j) / 2$, interaction strength $\epsilon_{ij}$, and interaction range $1/\alpha_{ij}$. The force $-\nabla V_\text{int}$ reaches its maximum value at $\ln 2/\alpha_{ij}$.
We define the total energy of a colloidal nanostructure as sum of pairwise Morse potentials between all bound spheres
\begin{equation}\label{eq:Etotsi}
    E(r) = \sum_{i\neq j} V_\text{int}(r_{ij}),
\end{equation}
where the reaction coordinate $r$ is defined as the distance between the arm ends.
As there are only two bonds involved to transition between the closed and the open configurations (\cf red particles in Fig.~\ref{fig:transition}), we modify the respective values $(\epsilon_\cc,\,\alpha_\cc)$ and $(\epsilon_\co,\,\alpha_\co)$ to get a desired target behavior.
We use a strong and short-ranged interaction between all permanently bound spheres using $\epsilon_\cb=100$ and $\alpha_\cb=50$ to prevent any other transitions. 

\begin{figure}[t]
    \includegraphics[width=0.49\textwidth]{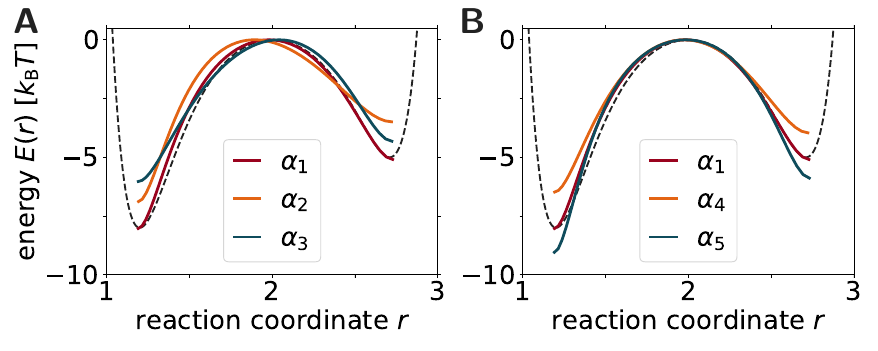}
    \caption{Effect of changing the inverse interaction ranges for closed and open bonds, $(\alpha_\cc,\alpha_\co)$, on the energy profiles, keeping permanent bonds and other parameters fixed. We use $\alpha_1=(9.12, 8.31)$ for reference and vary either the closed or open bond separately (\textsf{A}) with $\alpha_2=(9.12,5)$ and $\alpha_3=(6,8.31)$, or vary both by $\pm20\,\%$ (\textsf{B}) with $\alpha_4=(7.30,6.65)$ and $\alpha_5=(10.94,9.97)$. The fixed parameters are $\epsilon_\cc=10.67$, $\epsilon_\co=7.30$, and $R_\text{ref}=(3.51, -2.67, 1.26)$. The dashed line shows the target energy profile from Eq.~\eqref{eq:morse2} with parameters of the Machine nanostructure listed in Table~\ref{tab:params_target}.}
    \label{fig:energy_alphas}
\end{figure}

To gain intuition, Fig.~\ref{fig:energy_alphas} shows the effect of changing the inverse interaction ranges of the closed and open bonds, $(\alpha_\cc,\alpha_\co)$, on a given energy profile, while keeping all other model parameters fixed to $\epsilon_\cc=10.67$, $\epsilon_\co=7.30$, and $R_\text{ref}=(3.51, -2.67, 1.26)$.
Increasing both by $20\,\%$ leads to higher curvatures and larger energy barriers, whereas a decrease has the opposite effect with lower curvatures and smaller energy barriers. When only one of the interaction ranges is decreased the corresponding curvature is reduced as well as both energy barriers. Thus, we observe that the energy barriers change although we keep $\epsilon$ constant and only change the curvatures $\alpha$. This demonstrates that energy barriers and curvatures can not be controlled independently.

\subsection{Target energy profiles}
We model the target energy profiles of Machine (``easy'') and Source (``hard'') nanostructures through double-well potentials, allowing us to independently tune the positions of the closed, open, and transition states, all barrier heights, and the steepness of the barriers. Each energy profile is a continuous and differentiable combination of two Morse potentials
\begin{align}\label{eq:morse2}
    E(r) &= \epsilon_\cc \left(e^{-2\alpha_\cc (r - r_\cc)} - 2e^{-\alpha_\cc (r - r_\cc)}\right) + cr \nonumber\\ 
         &\quad + \epsilon_\co \left(e^{-2\alpha_\co (r_\co - r)} - 2e^{-\alpha_\co (r_\co - r)}\right).
\end{align}
The seven parameters include the locations of closed and open states, $r_\cc$ and $r_\co$, their respective binding energies, $\epsilon_\cc$ and $\epsilon_\co$, their inverse interaction ranges $\alpha_\cc$ and $\alpha_\co$ as well as a constant $c$. They are listed in Table~\ref{tab:params_target} for the target energy profiles of Machine and Source nanostructures.
\begin{table}[tb]\centering
  \begin{tabular}{cccccccc}
   \toprule
    & $r_\cc$ & $r_\co$ & $\epsilon_\cc$ & $\epsilon_\co$ & $\alpha_\cc$ & $\alpha_\co$ & $c$ \\ 
    \midrule
    Machine & $1.2$ & $2.71$ & $9.50$ & $6.43$ & $4.02$ & $3.94$ & $-0.04$ \\
    Source & $0.6$ & $3.75$ & $8.57$ & $34.16$ & $6.58$ & $1.09$ & $-0.05$ \\
   \bottomrule
  \end{tabular}
  \caption{List of all parameters modeling the target energy profiles of Machine (``easy'') and Source (``hard'') through double-well potentials as defined in Eq.~\eqref{eq:morse2}.}
  \label{tab:params_target}
\end{table}
Notice that the values of binding energies $\epsilon_\text{c/o}$ and inverse interaction ranges $\alpha_\text{c/o}$ differ for the target profiles and for the interaction potentials of bistable nanostructures, because the latter additionally include interactions of particles with permanent bonds in the total energy.

\subsection{Designing colloidal nanostructures with sphere-based arms}

We construct a sphere-based arm as a helical structure by mirroring one point of a tetrahedron with respect to the plane formed by the outermost three points, which is repeated until a desired number $N$ of colloids in the arm is reached. This stacking of regular tetrahedra is known as the Boerdijk–Coxeter helix~\cite{sado99}. Since we allow for variable diameters of spheres in the arm, we convert the bonds of the uniform helix to springs and run energy minimization on this spring network for variable sphere diameters in order to get the correct equilibrium distances. We mirror the arm and attach it to both sides of the hinge to get a symmetric structure.

\begin{figure}[t]
    \includegraphics[width=0.49\textwidth]{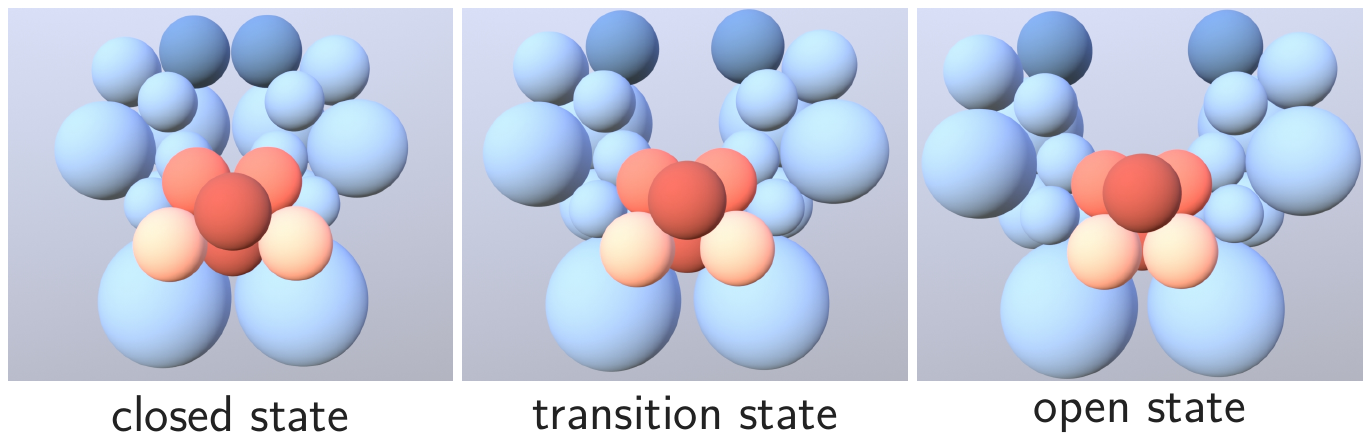}
    \caption{Bistable nanostructure transitioning from the polyhedral closed conformation through a transition state to the octahedral open conformation. The six-particle hinge structure is shown in red and the sphere-based arm model is shown in blue. In the closed state the dark red spheres in the center are bound, whereas in the open state the light red spheres are bound. The full DNEB animation of the nanostructure corresponding to the Machine in Scenario 4 is shown in the \href{https://seafile.ist.ac.at/f/fc768b2354374c2fb8d5/}{Supplemental video 1}.}
    \label{fig:transition}
\end{figure}

The \href{https://seafile.ist.ac.at/f/fc768b2354374c2fb8d5/}{Supplemental video 1} shows the transition pathway of a bistable nanostructure with sphere-based arms, corresponding to the Machine in Scenario 4. The hinge structure transitions from the polyhedral closed conformation through a transition state to the octahedral open conformation, as shown by the snapshots in Fig.~\ref{fig:transition}. 
We first calculate the energy profile of the hinge using the doubly-nudged elastic band (DNEB) method~\cite{tryg04} and refine the location of the transition state using eigenvector following~\cite{cerj81,wale04,maur05}. We construct the arms and add them to the hinge for all images of the transition pathway.

The optimized parameters used to model bistable nanostructures in all four design scenarios using the sphere-based arm model and the unconstrained arm model are listed in Table~\ref{tab:params_scns}.
\begin{table*}[t]
 \begin{tabular}{cccccccccc}
  \toprule
  \multicolumn{3}{c}{unconstrained arm design} & $\epsilon_\cb$ & $\epsilon_\cc$ & $\epsilon_\co$ & $\alpha_\cb$ & $\alpha_\cc$ & $\alpha_\co$ & $R_\text{ref}$ \\
  \midrule
  \multirow{4}{*}{\includegraphics[height=0.07\textwidth]{models1n}} & \multirow{2}{*}{Scenario 1} & Machine & $100$ & $11.54$ & $7.78$ & $50$ & $8$ & $8$ & $(3.35, -2.50, 1.09)$ \\
  & & Source & $100$ & $12.76$ & $35.79$ & $50$ & $8$ & $8$ & $(4.30, -4.73, -0.78)$ \\
  & \multirow{2}{*}{Scenario 3} & Machine & $100$ & $11.23$ & $7.54$ & $50$ & $8.16$ & $8.37$ & $(3.35, -2.50, 1.09)$ \\
  & & Source & $100$ & $9.09$ & $35.39$ & $50$ & $28.05$ & $4.87$ & $(4.96, -4.54, 1.15)$ \\
  \midrule
  \multicolumn{3}{c}{sphere-based arm design} & $\epsilon_\cb$ & $\epsilon_\cc$ & $\epsilon_\co$ & $\alpha_\cb$ & $\alpha_\cc$ & $\alpha_\co$ & diameters \\
  \midrule
  \multirow{4}{*}{\includegraphics[height=0.07\textwidth]{models2n}} & \multirow{2}{*}{Scenario 2} & Machine & $100$ & $11.54$ & $7.75$ & $50$ & $8$ & $8$ & $(0.32, 0.98, 1.00, 1.87, 2.89, 1.68, 3.08, 2.82, 1.44, 2.13, 1.18)$ \\
  & & Source & $100$ & $12.75$ & $35.79$ & $50$ & $8$ & $8$ & $(0.48, 1.03, 0.92, 1.97, 2.04, 1.41, 2.44, 1.25, 1.26, 1.49, 0.53)$ \\
  & \multirow{2}{*}{Scenario 4} & Machine & $100$ & $11.22$ & $7.53$ & $50$ & $8.16$ & $8.38$ & $(0.44, 1.17, 0.94, 1.08, 3.08, 1.28, 2.51, 2.63, 1.49, 2.06, 1.13)$ \\
  & & Source & $100$ & $10.04$ & $38.15$ & $50$ & $26.38$ & $4.20$ & $(0.46, 1.06, 0.94, 1.49, 2.69, 1.63, 7.72, 1.76, 1.36, 1.63, 0.54)$ \\
  \bottomrule
 \end{tabular}
 \caption{List of all parameters used to model bistable nanostructures with two different energy profiles in the four design scenarios. The values for interaction strengths and interaction ranges in the hinge correspond to the closed, open, and bound spheres. Scenarios 1 and 3 correspond to the unconstrained arm model and Scenarios 2 and 4 correspond to the sphere-based arm model.}
 \label{tab:params_scns}
\end{table*}

\subsection{Design freedom in the location of the transition state}

\begin{figure}[t]
    \includegraphics[width=0.5\textwidth]{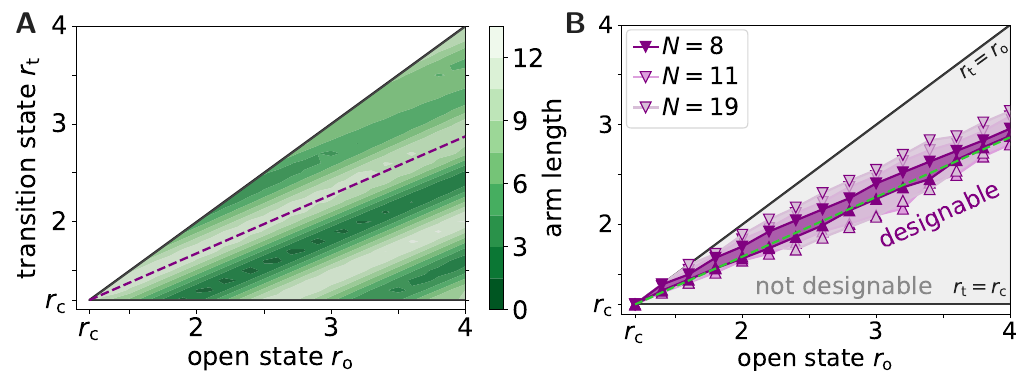}
    \caption{Design freedom in the location of the transition state in two design scenarios for modeling rigid arms. The locations of the meta-stable states are fixed at $r_\cc=1.2$ and $r_\co$. \textsf{(A)} For the unconstrained arm model (Scenario 1), all values of the transition state location $r_\ct$ between $r_\cc$ and $r_\co$ can be achieved with high accuracy, \ie, $\mathcal{L}_3^\mathrm{final}<10^{-12}$ for the loss from Eq.~(1) in the main text, using a spacing of 0.1. The color indicates the length of the arm, defined as distance from the binding site to the hinge center, in the optimal design. A dark green color means shorter distance, hence simpler arm design. \textsf{(B)} Control over the location of the transition state is limited in Scenario 2 when varying the colloid diameters in the sphere-based arm model. We find that the transition state is `nearly designable' (defined as $\mathcal{L}_3^\mathrm{final}<10^{-2}$) in the purple region, which increases for increasing numbers of colloids in the arms $N=8,11,19$, and is not designable in the gray region. The dashed lines can be calculated from the hinge model and match the design bands in both models. The comparison of these design scenarios informs how experimental constraints can impact functional design.}
    \label{fig:rt_opt}
\end{figure}

We know from previous work that arbitrary energy barriers can be designed with the hinge~\cite{good21}, but can any combination of $r_\cc$, $r_\co$, and $r_\ct$ be achieved just by changing the arm design? We test the design freedom in the location of the transition state for a fixed location of the closed state at $r_\cc=1.2$. The results for both arm models are shown in Fig.~\ref{fig:rt_opt}.
The comparison of these design scenarios informs how experimental constraints can impact functional design.

Importantly, the slope of the dashed lines matching the design bands can be calculated from the hinge model. The arms are attached to the hinge on planes formed by three hinge particles. The differences in angles between plane normals of the closed state ($\varphi_\cc$) and the transition state ($\varphi_\ct$) as well as the closed state and the open state ($\varphi_\co$) are given by
\begin{align}
    \varphi_\ct - \varphi_\cc \simeq 13.78, && \varphi_\co - \varphi_\cc \simeq 23.05.
\end{align}
The ratio of these two angle differences is 
\begin{equation}
    \frac{\varphi_\ct - \varphi_\cc}{\varphi_\co - \varphi_\cc} \simeq 0.598,
\end{equation}
and the dashed purple and green lines in Fig.~\ref{fig:rt_opt} obeying the relation
\begin{equation}
    r_\ct^\text{opt} = \frac{\varphi_\ct - \varphi_\cc}{\varphi_\co - \varphi_\cc} (r_\co - r_\cc) + r_\cc
\end{equation}
agree well with the slope of the designable region. This implies that robustly designable energy profiles are not symmetric, the transition state is biased towards the open state, due to our specific hinge design. 

\begin{figure}[t]
    \includegraphics[width=0.5\textwidth]{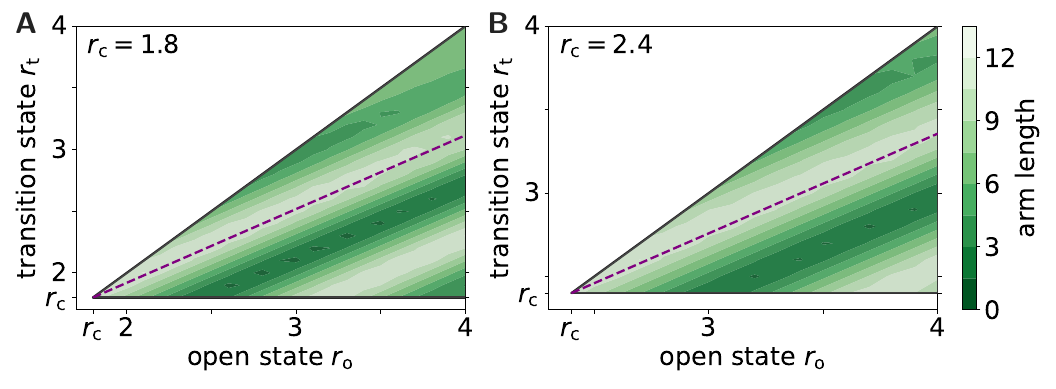}
    \caption{Design freedom in the location of the transition state with unconstrained arm design for fixed $r_\cc=1.8$ (\textsf{A}) and $r_\cc=2.4$ (\textsf{B}). This demonstrates that we can always exactly match the three locations of the minima and the transition state in the unconstrained arm model as the design is always successful, with the loss from Eq.~(1) in the main text converging to $\mathcal{L}_3(\theta)<10^{-12}$. For some combinations, however, at the cost of long arm lengths. The design bands are matched by the prediction from the hinge model as before.}
    \label{fig:rt_opt_rc}
\end{figure}

Figure~\ref{fig:rt_opt_rc} shows the design freedom for two higher values for the fixed location of the closed state, $r_\cc=1.8$ and $r_\cc=2.4$, in addition to the plot for $r_\cc=1.2$ shown in the main text and in Fig.~\ref{fig:rt_opt}.
The design with unconstrained arms is always successful, which demonstrates that we can always match target locations of both minima and the transition state as desired. We can thus tune all three locations precisely with unconstrained arms. For some combinations, however, at the cost of long arm lengths. The design bands are matched by the prediction from the hinge model as before.

\section{Navigating parameter space with differentiable programming}

\begin{figure}[t]
    \includegraphics[width=0.5\textwidth]{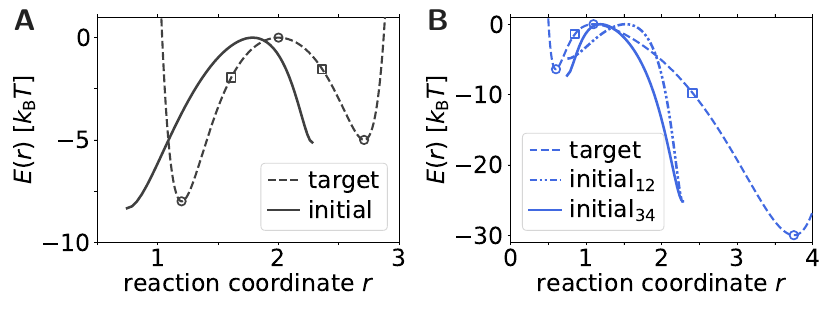}
    \caption{Energy profiles corresponding to the initial parameters used in the optimization presented in Figs.~2 and 4 in the manuscript to obtain target energy profiles. \textsf{(A)} For the easy profile (Machine), we use the same initial parameters, $\epsilon_\text{init}=(100,12,8)$ and $\alpha_\text{init}=(50,8,8)$ for (bound, closed, open) spheres, in all four scenarios. \textsf{(B)} For the hard profile (Source), we use $\epsilon_\text{init}=(100,10,30)$ and $\alpha_\text{init}^{12}=(50,8,8)$ in Scenarios 1 and 2, and an informed guess accounting for different slopes with $\alpha_\text{init}^{34}=(50,20,8)$ for variable interaction ranges in Scenarios 3 and 4.}
    \label{fig:Es_ini}
\end{figure}

We navigate parameter space with differentiable programming as described in Methods. 
The energy profiles corresponding to the initial parameters used in the optimization presented in Figs.~2 and 4 in the manuscript are shown in Fig.~\ref{fig:Es_ini} together with the target energy profiles.
For the arm design, we set all diameters to 1 in the sphere-based model, or use the corresponding location of the tip of the arm in the unconstrained model. For initial interaction energies and interaction ranges of the Machine profile, we use $\epsilon_\text{init}=(100,12,8)$ and $\alpha_\text{init}=(50,8,8)$ for (bound, closed, open) spheres in all four design scenarios. For the Source profile, we use $\epsilon_\text{init}=(100,10,30)$ and $\alpha_\text{init}^{12}=(50,8,8)$ in Scenarios 1 and 2, and an informed guess accounting for different slopes with $\alpha_\text{init}^{34}=(50,20,8)$ in Scenarios 3 and 4.

\begin{figure}[t]
    \includegraphics[width=0.49\textwidth]{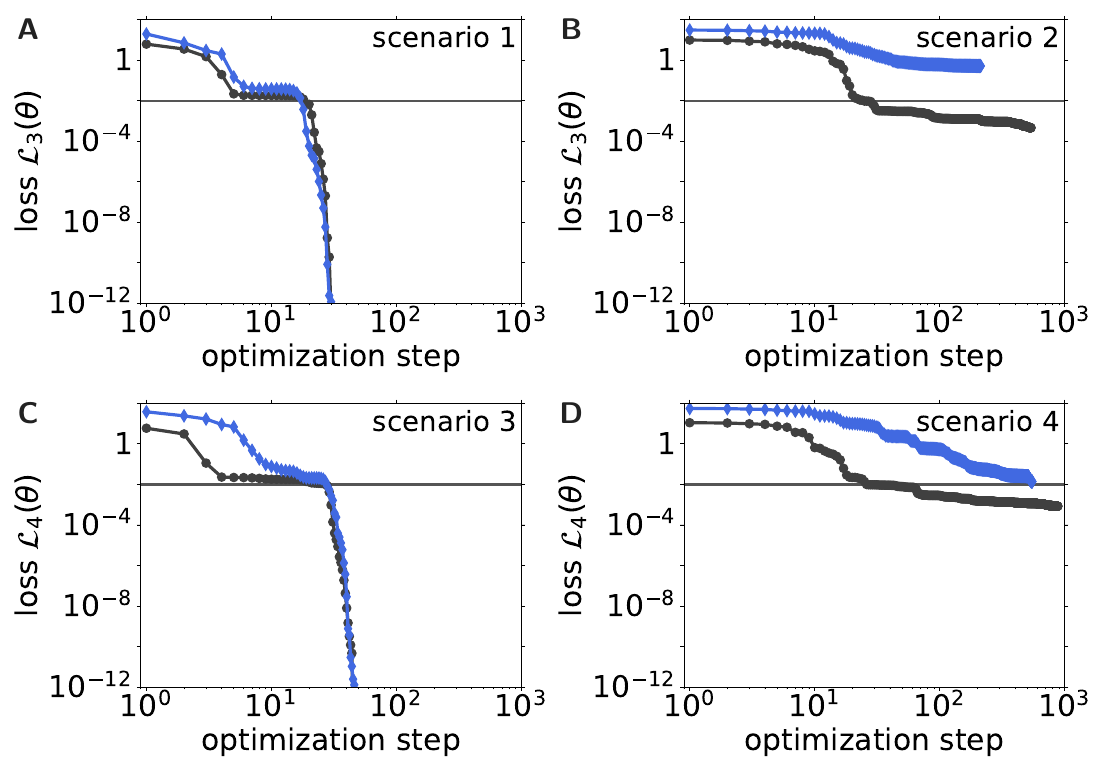}
    \caption{Loss function $\mathcal{L}(\theta)$ during optimization of the energy profiles in four different design scenarios for modeling bistable nanostructures. In Scenarios 1 and 2, the model parameters $\theta$ are given by the interaction strengths $\epsilon$ in the hinge as well as the relative location of the binding sites $R_\text{ref}$ of unconstrained arms \textsf{(A)} or the colloid diameters of sphere-based arms \textsf{(B)}. The model parameters in Scenarios 3 and 4, corresponding to panels \textsf{(C)} and \textsf{(D)} include the interaction ranges $1/\alpha$ additionally and the initial parameters are the optimal results of Scenarios 1 and 2, with constant interaction ranges, corresponding to energy profiles in Fig.~4\textsf{A-B} and Fig.~4\textsf{C-D} of the manuscript. The final, optimized parameters are used for the matched energy profiles presented in Fig.~4\textsf{E-F} and Fig.~4\textsf{G-H} of the manuscript.}
    \label{fig:loss}
\end{figure}

Figure~\ref{fig:loss} shows the loss function $\mathcal{L}(\theta)$ over the number of optimization steps for the target energy profiles in all four design scenarios. 
Our design framework works extremely well for nanostructures with unconstrained arms (Scenarios 1 and 3), which are precisely designable with final loss values below $10^{-12}$. These optimal loss values are significantly smaller for the unconstrained arm model compared to the sphere-based arm model, which are `nearly designable' with loss values about $10^{-3}$ for the Machine and $10^{-2}$ for the Source for variable interaction ranges. This agrees with the better match of target energy profiles in Fig.~4\textsf{E-F} compared to Fig.~4\textsf{G-H} of the manuscript. All final loss values are stated in Fig.~4 of the manuscript.

\begin{figure}[t]
    \includegraphics[width=0.49\textwidth]{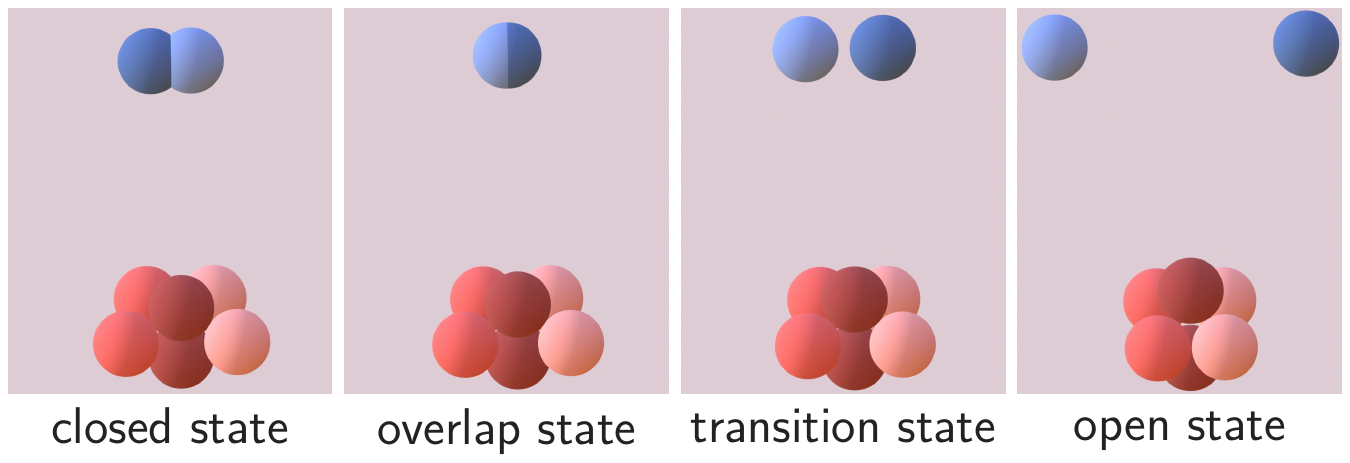}
    \caption{Source nanostructure with unconstrained arms optimized in Scenario 1 with loss 3 transitioning from the closed to the open state. We observe that the two arms ``pass through'' each other and the arm tips change sides during the transition (blue sphere versus light blue sphere). Note that we only show the ends of the unconstrained arms. The full DNEB animation is shown in the \href{https://seafile.ist.ac.at/f/8355f51b72184df6a18f/}{Supplemental video 2}.}
    \label{fig:transition_s1}
\end{figure}

A visual inspection of the optimal solution for Scenario 1 in Fig.~4\textsf{B} of the manuscript shows that the two arms ``pass through'' each other during the transition, as shown in Fig.~\ref{fig:transition_s1} by the blue spheres changing sides. It results in the reaction coordinate $r$ momentarily becoming zero. 
This is not inconsistent with $\mathcal{L}_3<10^{-12}$, because $\mathcal{L}_3$ does not depend on what happens in these intermediate regions; however, it implies that this solution is unphysical and highlights the importance of including additional physical constraints to prevent this.

\begin{figure}[t]
    \includegraphics[width=0.5\textwidth]{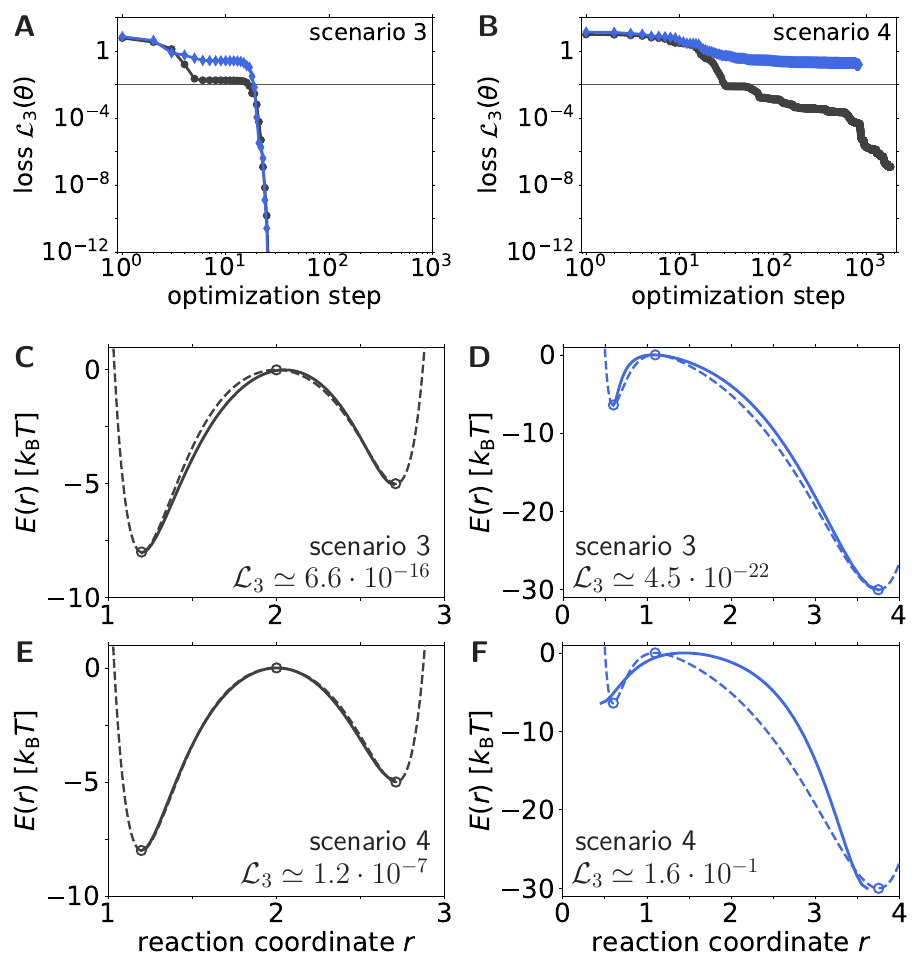}
    \caption{Optimization of the simpler loss function $\mathcal{L}_3(\theta)$ from Eq.~(1) in the manuscript for Scenarios 3 and 4. This works exceedingly well for Scenario 3 and motivates introducing a more ambitious loss function $\mathcal{L}_4(\theta)$.}
    \label{fig:loss1_scn34}
\end{figure}

Minimization of the simpler loss function $\mathcal{L}_3(\theta)$ from Eq.~(1) in the manuscript for Scenarios 3 and 4 works exceedingly well for Scenario 3 as shown in Fig.~\ref{fig:loss1_scn34}.
This motivates us to optimize the design space using the more ambitious loss function $\mathcal{L}_4(\theta)$ as presented in the manuscript. The discrepancies in Scenario 4 are not due to the constraints of the spheres. Rather, the problem stems from the indirect parameterization of the arms, which makes precise fine-tuning of the final $R_\mathrm{ref}$ difficult. Including energies at intermediate values helps solving this optimization problem as shown in Figs.~2 and 4 in the manuscript.

\subsection*{Hinge designs}

\begin{figure}[t]
    \includegraphics[width=0.49\textwidth]{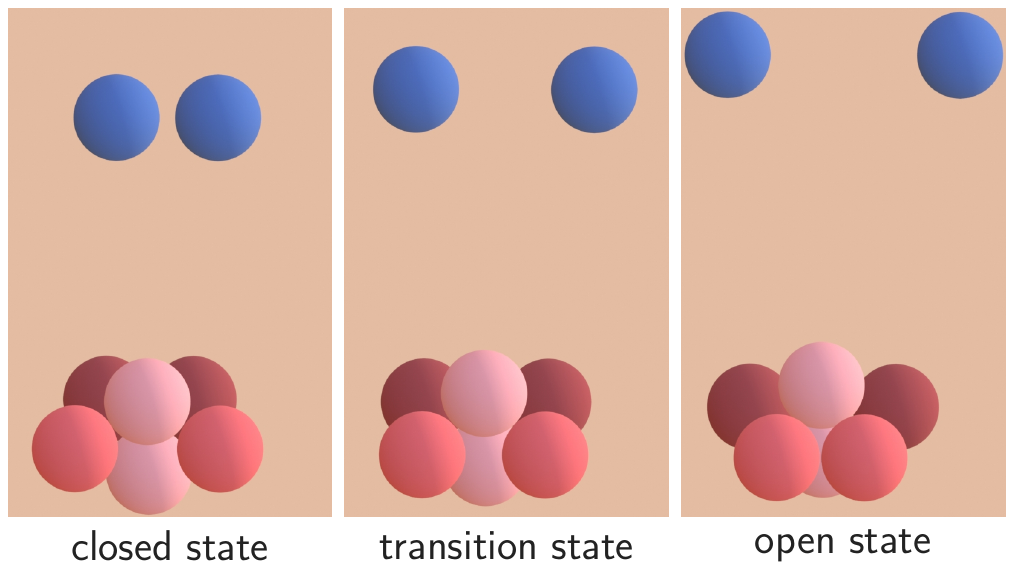}
    \caption{Bistable nanostructure consisting of a fully polyhedral hinge design and unconstrained arms. It transitions from a first polyhedral closed conformation, in which the dark red spheres in the back are bound, through a transition state to a second polyhedral open conformation, in which the red spheres in the front are bound. Note that we only show the ends of the unconstrained arms. The full DNEB animation is shown in the \href{https://seafile.ist.ac.at/f/c423a782970241f78c73/}{Supplemental video 3}.}
    \label{fig:hinge_trans}
\end{figure}

\begin{figure}[t]
    \includegraphics[width=0.5\textwidth]{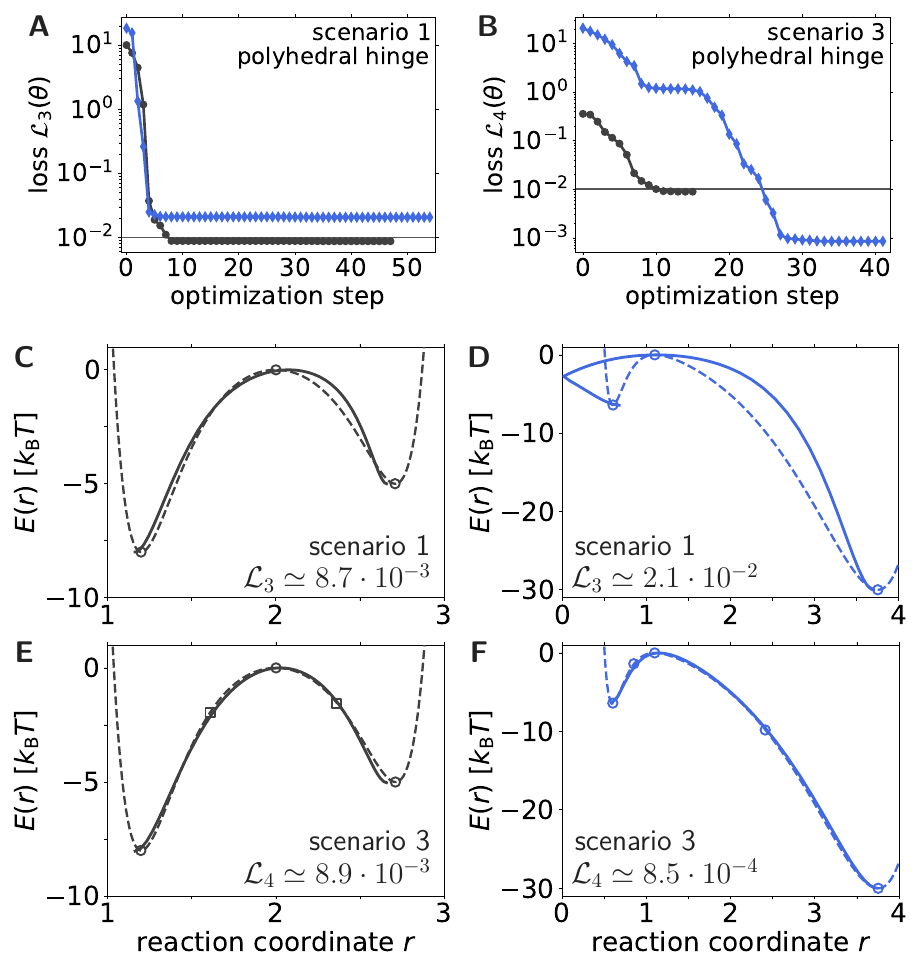}
    \caption{Optimization results for the fully polyhedral hinge design with unconstrained arms. The optimal designs in Scenario 3 show a good level of control and do not change qualitatively compared to those in the manuscript, suggesting that there are many possible hinge designs that yield similar target behavior.}
    \label{fig:hinge}
\end{figure}

We have tested other conformational transitions for the hinge structure, especially a hinge mechanism between two polyhedral conformations as shown in Fig.~\ref{fig:hinge_trans}. 
As in the previous hinge design, two pairs of particles are involved in the transition (red spheres in the front and dark red spheres in the back), but in this design, both  closed and open conformations are polyhedral.
We analyze design Scenarios 1 and 3 in the same way as in the manuscript for this new hinge design. The optimal designs in Scenario 3 show a good level of control as demonstrated in Fig.~\ref{fig:hinge} and do not change qualitatively compared to those in the manuscript. This suggests that there are many possible hinge designs that yield similar target behavior. It further presents evidence that our specific design allows for broader conclusions and qualitatively does not depend on the choice of the structural transition in the hinge.

\section{Functional speedup}

\begin{figure*}[htbp]
    \includegraphics[width=\textwidth]{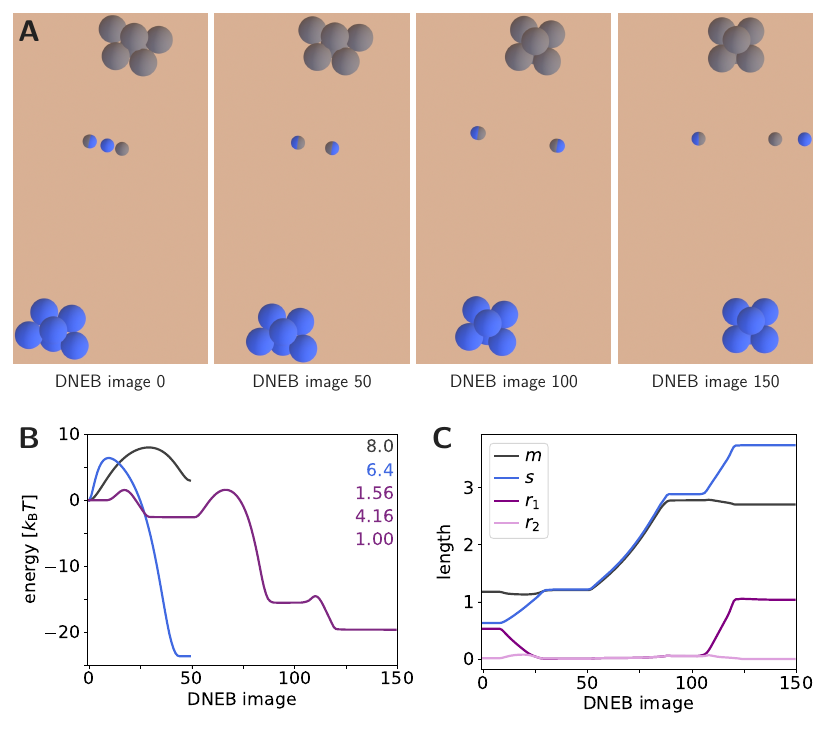}
    \caption{Analysis of the coupled energy-delivery reaction between a Machine and a Source nanostructure. \textsf{(A)} Snapshots of the coupled energy-delivery reaction between a Machine (gray) and a Source (blue) nanostructure for the optimized parameters in Scenario 3 for $\epsilon=10$. Initially, both nanostructures are in their closed polyhedral conformation and one binding site is bound. The Machine activates the Source until both binding sites are bound (DNEB image 50). Spontaneous fluctuations lead to forces exerted on the hinge structures. If these forces are large enough to overcome the energy barrier, both nanostructures open up together towards their octahedral conformation, corresponding to a long distance between both binding sites at DNEB image 100. Finally, one bond breaks due to the difference in open lengths between both nanostructures (DNEB image 150). Note that we only show the ends of the arms. An animation of the full transition pathway is shown in the \href{https://seafile.ist.ac.at/f/262de74bb8254227ade9/}{Supplemental video 4}. The transition pathway is obtained using the DNEB method~\cite{tryg04}. \textsf{(B)} The energy barriers along the coupled reaction (purple) are significantly smaller compared to the energy barriers of the Machine nanostructure (gray) transitioning on its own and the Source (blue) on its own. The exact values are given by the corresponding numbers. \textsf{(C)} Evolution of the four relevant lengths along the coupled reaction. First, the two nanostructures transition from their closed conformation to a meta-stable state with equal reaction coordinates $m$ and $s$. These lengths remain equal during the energy-delivery reaction where the separations of the binding sites ($r_1$ and $r_2$) are close to 0. Afterwards, one bond breaks to end up in the resulting open conformation.}
    \label{fig:speedup_analysis}
\end{figure*}

The \href{https://seafile.ist.ac.at/f/262de74bb8254227ade9/}{Supplemental video 4} shows an animation of the transition pathway of the coupled energy-delivery reaction, $\Mc\cdot\Sc \to \Mo\cdot\So$, of a Machine and a Source nanostructure for the optimized parameters in Scenario 3 for $\epsilon=10$. Four snapshots are presented in Fig.~\ref{fig:speedup_analysis} to illustrate the transition pathway along the coupled reaction obtained using the DNEB method~\cite{tryg04}.
Initially, both nanostructures are in their closed polyhedral conformation and one binding site is bound. The Machine activates the Source until both binding sites are bound, as shown in DNEB image 50. Spontaneous fluctuations lead to forces exerted on the hinge structures. If these forces are large enough to overcome the energy barrier, both nanostructures open up together towards their octahedral conformation, corresponding to a long distance between both binding sites (DNEB image 100). Finally, one bond breaks due to the difference in open lengths between both nanostructures to result in DNEB image 150. 
Figure~\ref{fig:speedup_analysis}\textsf{B} demonstrates that the energy barriers along the coupled reaction, $\Mc\cdot\Sc \to \Mo\cdot\So$, (purple curve) are significantly smaller compared to the energy barriers of the Machine nanostructure transitioning on its own, $\Mc\to\Mo$, and the Source on its own, $\Sc\to\So$. Comparing the mean first-passage time of the coupled reactions to the Machine on its own results in the functional speedup presented in the manuscript.
The four relevant lengths along the coupled reaction are plotted in Fig.~\ref{fig:speedup_analysis}\textsf{C}. First, the two nanostructures transition from their closed conformation to a meta-stable state with equal reaction coordinates $m$ and $s$. These lengths remain equal during the energy-delivery reaction where the separations of the binding sites are close to 0. Afterwards, one bond breaks to end up in the resulting open conformation.

\begin{figure}[t]
    \includegraphics[width=0.49\textwidth]{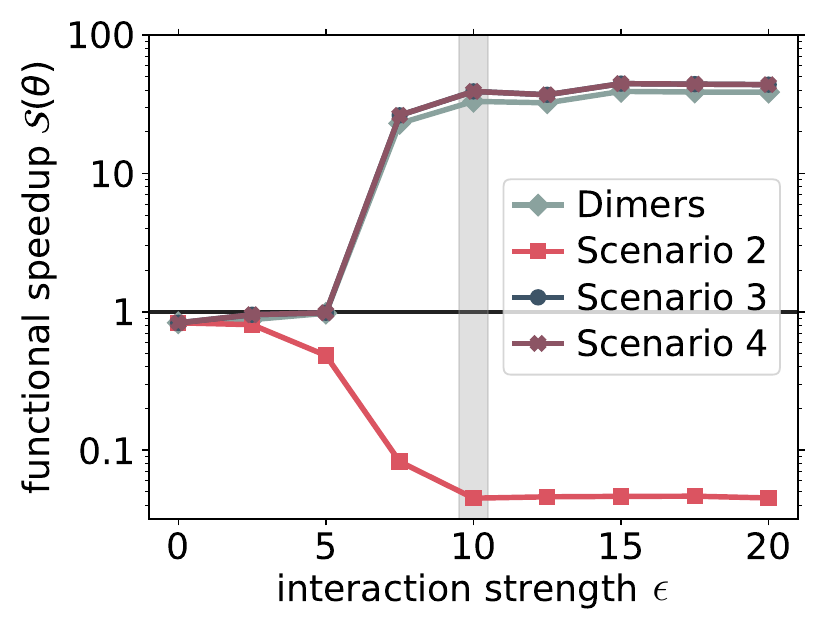}
    \caption{Functional speedup as function of the interaction strength $\epsilon$. The data shows that for interaction strengths $\epsilon>5\,\kT$, the energy-delivery mechanism works and the mutual transition of Machine and Source nanostructures is significantly sped up for Scenarios 3 and 4, comparably to the dimer reference, whereas the Source is actually harmful in Scenario 2. The gray bar marks the value $\epsilon=10\,\kT$ for which the data is plotted in Fig.~5 in the main text.}
    \label{fig:speedup_eps}
\end{figure}

The functional speedup as function of the interaction strength $\epsilon$ is shown in Fig.~\ref{fig:speedup_eps}.
The data shows that for interaction strengths $\epsilon>5\,\kT$, the energy-delivery mechanism works and the mutual transition of Machine and Source nanostructures is significantly faster for Scenarios 3 and 4, comparably to the dimer reference, whereas the Source is actually harmful in Scenario 2.

\section{Supplemental Media}

\href{https://seafile.ist.ac.at/f/fc768b2354374c2fb8d5/}{Supplemental video 1}
Video showing the transition pathway of a bistable nanostructure with sphere-based arms, corresponding to the Machine in Scenario 4.

\href{https://seafile.ist.ac.at/f/8355f51b72184df6a18f/}{Supplemental video 2}
Video showing the transition pathway of the Source in Scenario 1. The arm tips change sides during the transition, demonstrating that the arms pass through each other.

\href{https://seafile.ist.ac.at/f/c423a782970241f78c73/}{Supplemental video 3}
Video showing the transition pathway of a fully polyhedral hinge structure with unconstrained arms. Note that we only show the ends of the arms.

\href{https://seafile.ist.ac.at/f/262de74bb8254227ade9/}{Supplemental video 4} 
Video showing the transition pathway of the coupled energy-delivery reaction of a Machine (gray) and a Source (blue) nanostructure for the optimized parameters in Scenario 3. Note that we only show the ends of the arms.

\end{document}